\documentclass[aps,prd]{revtex4}
\usepackage{epsfig,epsf}
\usepackage{amsmath}
\usepackage{amsthm}
\usepackage{amsfonts}
\usepackage{amssymb}
\usepackage{dsfont}
\usepackage{multirow}
\usepackage{appendix}
\usepackage{slashed}
\usepackage[active]{srcltx}
\usepackage{psfrag}
\usepackage{subfigure}
\usepackage[colorlinks,citecolor=blue,urlcolor=red,linkcolor=red]{hyperref}

\begin{document}
\title{{\Large{\bf Analysis of the semileptonic $B\to K_1 \ell^+
\ell^-$transitions and non-leptonic $B \to K_1 \gamma$ decay in the
AdS/QCD correspondence }}}

\author{\small
S. Momeni\footnote {e-mail: samira.momeni@phy.iut.ac.ir}, R.
Khosravi\footnote {e-mail: rezakhosravi @ cc.iut.ac.ir}}

\affiliation{Department of Physics, Isfahan University of
Technology, Isfahan 84156-83111, Iran }

\begin{abstract}
We consider the axial-vector mesons $K_1(1270)$ and $K_1(1400)$ as a
mixture of two $|^3P_1\rangle$ and $|^1P_1\rangle$ states with the
mixing angle $\theta$ that equal to $(-34\pm 13)^\circ$. We
calculate the light-front distribution amplitudes (LFDAs) and  decay
constant formulas for both the axial-vector mesons $K_1$ in the
AdS/QCD correspondence. The transition form factors of the
semileptonic $B\to K_1$ decays are derived in terms of the LFDAs for
$K_1$ mesons. Using these form factors and decay constant values,
the differential branching ratios  of $B\to K_1 (1270, 1400) \ell^+
\ell^-$,~$\ell=\mu,\tau$ transitions are plotted with respect to the
four-momentum transfer squared, $q^2$. In addition, the branching
ratio values of these decays and the non-leptonic $B \to K_1(1270,
1400) \gamma$ decays are estimated. A comparison is made between our
results for the branching ratios of $B\to K_1(1270, 1400) \gamma$
decays in the AdS/QCD model and predictions obtained from the
light-cone sum rules (LCSR) as well as the experimental values.
Finally, the forward-backward asymmetries for the aforementioned
semileptonic decays are plotted on $q^2$ in both the AdS/QCD
correspondence and two Higgs doublet model (2HDM) in order to test
the standard model (SM) and search for the new physics (NP).

\end{abstract}

\pacs{11.15.Tk, 11.25.Tq, 13.20.He, 14.40.Df}

\maketitle

\section{Introduction}\label{sec.1}

Inclusive and exclusive decays of $B$ meson improve our studies in
understanding the dynamics of quantum chromo dynamics (QCD). Among
of all $B$ decays, the theoretical description of the semileptonic
decays is relatively simple. These semileptonic decays usually occur
by two various diagrams: a) simple tree diagrams which can be
performed via the weak interaction, b) electroweak penguin and  box
diagrams which can be fulfilled through the flavor changing neutral
current (FCNC) transitions in the SM. The FCNC decays $B \to K_1
\ell^+ \ell^- $, involving the axial-vector strange mesons, have
been the subjects of many theoretical studies, since they are
important for a few reasons. They are sensitive to NP contributions
to penguin operators. Therefore,  we can check the SM and search NP
by estimating the SM predictions for these decays and comparing
these results to the corresponding values from some NP models. On
the other hand, in particle physics, reliable calculations of
heavy-to-light transition form factors of semileptonic $B$ decays
are very important since they are also used to determine the
amplitude of non-leptonic $B$ decays applied to evaluate the CKM
parameters as well as to test various properties of the SM.

Sofar, the heavy-to-light transitions $B \to K_1 \ell^+ \ell^- $,
as a FCNC process, have been studied in many theoretical approaches
in the frame work of the SM such as the three-point QCD sum rules
(3PSR) \cite{Dag,Bayar}, the LCSR \cite{Y1,Yang1,MoKh}, perturbative
QCD (PQCD) approach \cite{Li,Li2} and light-front quark model (LFQM)
\cite{Cheng,Verma}; and some NP models, such as universal extra
dimension \cite{Ahmed,Saddique,Paracha}, models involving
supersymmetry \cite{Bashiry}, the fourth-generation fermions
\cite{Rehman}, the 2HDM \cite{Falahati}, the non-universal $Z'$
model \cite{Hua} and the model-independent new-physics corrections
to the Wilson coefficients \cite{Hatanaka}. Considering the physical
observables of these decays, such as the branching ratio value,
dilepton invariant mass spectrum, forward-backward asymmetry and
double lepton polarization provide us a lot of useful information.
In this paper, we plan to investigate the FCNC $B \to K_1$
transitions in the AdS/QCD correspondence.

The interactions among quarks and gluons, described by QCD, are
particularly important because they exhibit many characteristic and
challenging features of a strongly-coupled theory. In the high
momentum transfer regime, QCD is asymptotically free and can be
considered with methods of perturbation theory. In the low momentum
transfer regime, confinement is created and QCD becomes
strongly-coupled. Therefore, one of the most important issues of
strong interaction dynamics is to obtain analytic solutions for the
wave functions of hadrons outside of the perturbative regime. One of
the proposed ideas for overcoming these problems is based on the
light-front QCD and using the AdS/CFT correspondence
\cite{JMMald1,JMMald2} between string states in anti-de Sitter (AdS)
space and conformal field theories (CFT) in physical space-time
\cite{BroTera1,BroTera2,BroTera3,BroTera4,BroTera5,BroTera6}. The
application of the AdS space and conformal methods to QCD can be
motivated from the experimental evidence \cite{ADeVBu}, and
theoretical discussions that the QCD coupling $\alpha_s(Q^2)$ has an
infrared fixed point at low $Q^2$ \cite{SJBroRSh,GFdeSJBr}. In this
region, the AdS/QCD approach has been successful in obtaining
general properties of phenomenological QCD such as hadronic spectra,
decay constants, and wave functions
\cite{Grigoryan1,Grigorya,Hong1,Radyushkin}.

There is a significant mapping between the AdS space  description of
hadrons and the light-front wave functions (LFWFs) of bound states
in QCD quantized on the light-front, known as holographic LFWFs (for
instance see \cite{BroTera6}). The LFWFs in QCD, similar to the
Schrodinger wave functions of atomic physics, provide an explanation
of the structure and internal dynamics of hadrons in terms of their
constituent quarks and gluons. However, they are determined at fixed
light-front time instead of at fixed ordinary time \cite{GFdeSJBr}.
Using the LFWF, some physical quantities related to hard exclusive
reactions can be calculated such as distribution amplitudes,  form
factors and structure functions.

The holographic LFWF  has been successfully applied  to describe
diffractive $\rho$ meson electroproduction at HERA \cite{Forshaw}.
In addition, this LFWF has been used to study the spectrum
\cite{Branz} and  the distribution amplitudes (DAs) of light and
heavy mesons \cite{Hwang12}. After introducing the light-front
spinor structure of the wave functions for light vector mesons in
analogy with that of the photon, light-front distribution amplitudes
(LFDAs) of the $\rho$ and $K^*$ vector mesons have been predicted in
$B \to \rho \gamma$ \cite{Ahma1}, and $B \to K^* \gamma$
\cite{Ahma2} decays. Also, using the holographic DAs, the transition
form factors of the semileptonic $B \to \rho$ \cite{Ahma3}, and $B
\to K^*$ decays \cite{Ahma4} have been estimated. These form factors
have been then utilized to make predictions for the isospin
asymmetry of $B \to K^* \mu^+ \mu^-$ transition \cite{Ahma5} and for
branching ratio values of the semileptonic $B \to \rho \ell \nu$
decays \cite{AhmaLord}. Dynamical spin effects have been taken into
account of the holographic pion wave function in order to predict
its mean charge radius, decay constant, space-like electromagnetic
form factor, twist-2 DA and the photon-to-pion transition form
factor \cite{AhmaChish}. Recently, the AdS/QCD DAs of pseudoscalar
mesons and their application to $B$-meson decays have been studied
in Ref. \cite{ChaBro,MomenKhosravi}.

Sofar, the holographic DAs  have been not calculated for
axial-vector mesons. The study of the DAs for axial-vector mesons is
important for considering exclusive decays such as $B\to K_1(1270)
\gamma$.  The branching ratio value of the aforementioned decay has
been measured by Belle \cite{HYang}, whereas the axial-vector meson
$K_1(1270)$ is a mixtures of two $|^3P_1\rangle$ and $|^1P_1\rangle$
states. Usually, the DAs for light mesons are estimated from the
LCSR method, known as light-cone distribution amplitudes (LCDAs). In
this work, we plan to calculate the holographic DAs and tensor decay
constants for the axial-vector mesons $K_1(1270)$ and $K_1(1400)$.
Due to the axial-vector masons $K_1$ are  considered as a mixture of
two states, we need to investigate the holographic DAs for
$|K_{1A}\rangle$ and $|K_{1B}\rangle$ states in the AdS/QCD
correspondence in terms of the LFWFs.  Then,  we can derive the DAs
for $K_1$ mesons in terms of the holographic DAs for these states.
Inserting the holographic DAs for $K_1$ in the transition form
factor equations of the semileptonic $B \to K_1$ decays, which have
been calculated via the LCSR method \cite{MoKh}, we can predict the
branching ratio value for  $B \to K_1(1270) \gamma$ decay.

The main purpose of this work is as follows:

$\bullet$  Investigation of the holographic DAs for  the
axial-vector mesons $K_1(1270)$ and $K_1(1400)$ in the AdS/QCD
correspondence. It would be reminded that an accurate calculation of
the DAs is very important since they provide a major source of
uncertainty in the theoretical predictions of the physical
quantities.

$\bullet$ Calculation of  the  tensor decay constants for the
axial-vector mesons $K_1(1270,1400)$  and considering the form
factors of  $B \to K_1(1270, 1400) \ell^+ \ell^-$ decays in order to
investigation the dilepton invariant mass spectrums and prediction
of the branching ratio values of them.

$\bullet$ Predictions of the branching ratio values for the
non-leptonic $B \to K_1(1270, 1400) \gamma$ decays. A comparison is
made between our result for $B \to K_1(1270) \gamma$ decay and the
experimental value.

$\bullet$ Considering  the forward-backward asymmetries for  $B \to
K_1(1270, 1400) \ell^+ \ell^-$ transitions on $q^2$ in the AdS/QCD
correspondence and 2HDM in order to test the SM and search for the
NP.

The contents of this paper are as follows: In section II, the LFWFs
for the axial-vector mesons $K_1(1270,1400)$ are calculated in the
frame work of the AdS/QCD. Then, the decay constant formulas and
LFDAs for $K_1$ are derived. For this purpose, we investigate the
holographic DAs for $|K_{1A}\rangle$ and $|K_{1B}\rangle$ states in
the AdS/QCD correspondence in terms of the LFWFs. In section III, we
analyze the LFDAs and decay constants for $K_1$ mesons and compare
our results with predictions of the LCSR method. Applying the LFDAs
of $K_1$ mesons in the transition form factors of the FCNC $B\to
K_1$ decays, we analyze these form factors as well as the dilepton
invariant mass spectrum on $q^2$. In addition, we obtain the
branching ratio values for  $B\to K_1(1270,1400) \ell^{+} \ell^{-}$
and $B\to K_1(1270,1400) \gamma$ decays. Our result for the
branching ratio of the non-leptonic decay $B\to K_1(1270) \gamma$ is
compared with the experimental value. Finally, the forward-backward
asymmetries for $B \to K_1(1270, 1400) \ell^+ \ell^-$ transitions,
with respect to $q^2$, are compared in the AdS/QCD correspondence
and 2HDM.

\section{Distribution amplitudes and decay constants in AdS/QCD}\label{sec.2}

The physical states of $K_1(1270)$ and $K_1(1400)$ mesons are
considered as a mixture of two $|^3P_1\rangle$ and $|^1P_1\rangle$
states and can be parameterized in terms of a mixing angle
$\theta_K$, as follows:
\begin{eqnarray}\label{eq21}
|K_1(1270)\rangle &=&\sin\theta_K |^3P_1\rangle +
\cos\theta_K |^1P_1\rangle,\nonumber\\
|K_1(1400)\rangle &=&\cos\theta_K |^3P_1\rangle - \sin\theta_K
|^1P_1\rangle,
\end{eqnarray}
where $|^3P_1\rangle\equiv |K_{1A}\rangle$ and $|^1P_1\rangle\equiv
|K_{1B}\rangle$ have different masses and decay constants. Also, the
mixing angle $\theta_K$  can be determined by the experimental data.
There are various approaches to estimate the mixing angle. The
result $35^\circ \leq |\theta_K| \leq 55^\circ$ was found in Ref.
\cite{Burakovsky}, while two possible solutions were obtained as
$|\theta_K|\approx 33^\circ \vee 57^\circ$  in Ref. \cite{Suzuki}
and as $|\theta_K|\approx 37^\circ \vee 58^\circ$ in Ref.
\cite{HYCheng}. A new window for the value of $\theta_K$ is
estimated from the result of $\tau \to ¨ K_1(1270)\nu_{\tau}$ data
as \cite{Hatanaka2}
\begin{eqnarray}
\theta_K = {-(34\pm13)}^{\circ}.
\end{eqnarray}
Sofar this value is used in Refs.
\cite{Dag,Bayar,Yang1,Bashiry,Falahati,Hatanaka}. In this study, we
also use the result of $\theta_K = {-(34\pm13)}^{\circ}$.

The twist-2 DAs,  $\Phi_{K_1}^{\parallel,\perp}$, for $K_1$ mesons
are given in terms of the twist-2 DAs of $K_{1A}$ and $K_{1B}$
states, $\Phi^{\parallel,\perp}_{K_{1A}}(u)$ and
$\Phi^{\parallel,\perp}_{K_{1B}}(u)$, as \cite{Y1}:
\begin{eqnarray}\label{eq22}
\Phi^\parallel_{K_1}(u) &=& C_1\, \frac{f_{K_{1A}}
m_{K_{1A}}}{f_{K_1} m_{K_1}} \Phi^\parallel_{K_{1A}}(u) + C_2\,
\frac{f_{K_{1B}} m_{K_{1B}}}{f_{K_{1}} m_{K_1}}
\Phi^\parallel_{K_{1B}}(u), ~~~~\nonumber\\
\Phi^\perp_{K_1}(u) &=&C_1\, \frac{f_{K_{1A}}^\perp}{f_{K_1}^\perp}
\Phi^\perp_{K_{1A}}(u) + C_2\, \frac{f_{K_{1B}}^\perp
}{f_{K_{1}}^\perp} \Phi^\perp_{K_{1B}}(u),
\end{eqnarray}
where $(C_1,C_2)=(\sin{\theta_K},\cos{\theta_K})$ for $K_1(1270)$
meson, and $(C_1,C_2)=(\cos{\theta_K},-\sin{\theta_K})$ for
$K_1(1400)$. In this phrases, $u$ refer to the momentum fraction
carried by the quark in $K_{1}$. In addition, $f_{K_{1}}$ and $
f_{K_{1}}^{\perp}$ are  decay constants, written in terms of
$f_{K_{1A(1B)}}$ and $f^{\perp}_{K_{1A(1B)}}$ as
\begin{eqnarray}\label{eq23}
f_{K_1} &=&C_1\, \frac{ m_{K_{1A}}}{ m_{K_1(1270)}} f_{K_{1A}}+
C_2\,\frac{ m_{K_{1B}}}{ m_{K_1(1270)}}
a_0^{\parallel,K_{1B}} f_{K_{1B}}, ~~~~\nonumber\\
f_{K_1}^{\perp} &=& C_1\, a_0^{\perp,K_{1A}} f_{K_{1A}}^{\perp}+
C_2\,f_{K_{1B}}^{\perp},
\end{eqnarray}
where $a_0^{\perp,K_{1A}}$ and $a_0^{\parallel,K_{1B}}$ are G-parity
invariant Gegenbauer moments for $K_{1A}$ and $K_{1B}$ states which
have been estimated in Ref. \cite{Y1}.

First, we aim to calculate the twist-2 DAs for $K_1$ mesons in the
AdS/QCD correspondence. According to Eq. (\ref{eq22}), we need to
investigate the twist-2 DAs for two states $K_{1A}$ and $K_{1B}$ in
terms of the holographic LFWFs.  In order to consider the twist-2
DAs, the matrix elements of $K_{1A}$ and $K_{1B}$ states should be
considered. For instance, the following two-particle matrix elements
of state $K_{1A}$ in the light-front coordinate, $x^{\mu}=(x^+, x^-,
\textbf{x}_{\perp})$, at equal light-front time $x^+$, are written
as:
\begin{eqnarray}
\langle 0|\bar u(0) \gamma^\mu \gamma_5 s(x^{_-})| K_{1A}(p, \lambda
)\rangle =  f_{K_{1A}} m_{K_{1A}} \int_0^1 du \; e^{-iu p^+x^-} \{
p^{\mu} \frac{\varepsilon_{\lambda}. x}{p^{+}\, x^{-}}
\Phi^\parallel_{K_{1A}}(u,\mu) +\cdots \},\label{eq24}\\
\langle 0|\bar u(0) \sigma^{\mu\nu}\gamma_{5} s(x^{_-}) |K_{1A}(p,
\lambda )\rangle  = i f_{K_{1A}}^{\perp} \int_0^1 du \, e^{-i u
p^{+} x^{-}} \{(\varepsilon^{\mu}_{\lambda} p^{\nu} -
\varepsilon^{\nu}_{\lambda} p^{\mu})\, \Phi^\perp_{K_{1A}}(u,\mu) +
\cdots\},\label{eq25}
\end{eqnarray}
where $\gamma^\mu=(\gamma^+, \gamma^-, \gamma^1, \gamma^2)$.  The
$"\cdots"$ describes the contributions coming from higher twist DAs.
In these relations, $p^+$ is the "plus" component of the
four-momentum of  $K_{1A}$ state  given by
$p^{\mu}=\left(p^{+},\frac{m_{K_{1A}}^2}{p^{+}} ,0_{\perp} \right)$.
The polarization vectors $\varepsilon_\lambda~ (\lambda=L, T)$ for
state $K_{1A}$ are chosen as $
\varepsilon_{L}=\left(\frac{p^{+}}{m_{K_{1A}}},\frac{m_{K_{1A}}}{p^{+}},0_{\perp}\right)$,
and $\varepsilon_{T(\pm)}=\frac{1}{\sqrt{2}}\left(0,0, 1, \pm i
\right) $.

Taking $\lambda=L$ and $\mu=+$  in Eq. (\ref{eq24}), in addition,
the scalar product of Eq. (\ref{eq25}) in
$(\varepsilon^{*}_{T})_{\mu}$, we obtain:
\begin{eqnarray}
\langle 0|\bar u(0) \gamma^{+}\gamma_{5} s(x^-)|K_{1A} (p,L) \rangle
=  f_{K_{1A}} m_{K_{1A}} \int_0^1 du \; e^{-iu p^+x^-} [ p^{+}
(\frac{\varepsilon_{L}. x}{p^{+} x^{-}})\,
\Phi^\parallel_{K_{1A}}(u,\mu) ],\label{eq26}\\
\langle 0|\bar u(0)
[\gamma.\varepsilon^{*}_{T},\gamma^{+}]\gamma_{5} s(x^-)|K_{1A}
(p,\pm) \rangle  = 2\,f_{K_{1A}}^{\perp}  \int_0^1 du \, e^{-i u
p^{+} x^{-}} [p^{+}(\varepsilon^{1}_{T}\mp i\varepsilon^{2}_{T})
\Phi^\perp_{K_{1A}}(u,\mu)],\label{eq27}
\end{eqnarray}
where $\gamma.\varepsilon^{*}_{T}$ is placed instead of $\gamma_1\mp
i \gamma_2 $. Applying the Fourier transform of the above matrix
elements with respect to the longitudinal distance $x^{-}$, the
twist-2 DAs are given by:
\begin{eqnarray}
\Phi_{K_{1A}}^{\parallel}(\alpha,\mu)&=& \frac{1}{f_{K_{1A}}}\int
dx^- \; e^{i \alpha p^+ x^-}\langle 0|\bar u(0)
\gamma^{+}\,\gamma_{5} s(x^-)|K_{1A}
(p,L) \rangle,\label{eq28}\\
\Phi_{K_{1A}}^{\perp}(\alpha,\mu)&=& \frac{1} {2 f_{K_{1A}}^{\perp}}
\int  dx^- \; e^{i \alpha p^+ x^-} \langle 0|\bar u(0)
[\gamma.\varepsilon^{*}_{T},\gamma^{+} ]\,\gamma_{5} s(x^-)|K_{1A}
(p,\pm) \rangle,\label{eq29}
\end{eqnarray}
where $\alpha$ is the momentum fraction of quark in state $K_{1A}$.

To obtain $\Phi_{K_{1A}}^{\parallel, \perp}$  in  Eqs. (\ref{eq28})
and (\ref{eq29}), we should calculate the matrix elements which
appear  in these relations. These matrix elements can be estimated
by using the LFWF, $\Psi^{K_{1A},\lambda}_{h,\bar{h}}(\alpha,
\mathbf{k})$ of the $K_{1A}$ state as \cite{R1}:
\begin{eqnarray}\label{eq210}
p^+\int dx^- e^{i\alpha p^+x^-} \langle 0 | \bar{u}(0)  \Gamma
s(x^-) |K_{1A}(p, \lambda) \rangle &=& \sqrt{\frac {N_c}{4\pi}}
\sum_{h,\bar{h}} \int^{|\mathbf{k}| < \mu}
\frac{d^{2}\textbf{k}}{(2\pi)^2} \;
\Psi^{K_{1A},\lambda}_{h,\bar{h}}(\alpha,\mathbf{k}) \nonumber \\
&\times& \left \{
\frac{\bar{v}_{\bar{h}}((1-\alpha)p^{+},-\mathbf{k})}{\sqrt{(1-\alpha)}}
\Gamma \frac{u_h(\alpha \,p^+,\mathbf{k})}{\sqrt{\alpha}} \right \},
\end{eqnarray}
while $\Gamma$ stands for $\gamma^+\gamma_5$ and
$[\gamma.\varepsilon^{*}_{T}, \gamma^+]\gamma_5$.  Here $\mathbf{k}$
is transverse momenta of quark, and the renormalization scale $\mu$
is identified with the ultraviolet cut-off on $\mathbf{k}$
\cite{Kogut,Diehl}. Also,  $ u (\bar{v})$ and $h (\bar{h})$ are the
spinor and helicity of quark (anti-quark), respectively.  The
explicit expressions for light-front spinors with positive and
negative helicities have been given in Ref. \cite{Lepage1}. Using
these expressions for the light-front spinors $\bar{v}_{\bar h}$ and
$u_{h}$, We obtain:
\begin{eqnarray}
\frac{\bar{v}_{\bar{h}}}{\sqrt{(1-\alpha)}}\gamma^{+}\,\gamma^{5}\frac{u_{h}}{\sqrt{\alpha}}
&=&
p^{+}\,(\delta_{\bar{h}{+},h{-}}-\delta_{\bar{h}{-},h{+}}),\label{eq211}\\
\frac{\bar{v}_{\bar{h}}}{\sqrt{(1-\alpha)}}
[\varepsilon^{*}_{\pm}.\gamma,\gamma^+]\,\gamma^{5}
\frac{u_h}{\sqrt{\alpha}} &=& \mp 4\sqrt{2} p^+ \delta_{h\pm,\bar{h}
\pm}\, ,\label{eq212}
\end{eqnarray}
where  $h +$ and $h-$ are  used for  positive and  negative
helicity, respectively. The LFWF of $K_{1A}$ in Eq. (\ref{eq210}) is
defined in momenta space as \cite{R1}:
\begin{equation}\label{eq213}
\Psi^{K_{1A},\lambda}_{h,\bar{h}}(\alpha,\mathbf{k})=\sqrt{\frac{N_c}{4\pi}}
S_{h,\bar{h}}^{K_{1A},\lambda}(\alpha,\mathbf{k})\,
\phi^{K_{1A}}_{\lambda}(\alpha,\mathbf{k}).
\end{equation}
In Refs. \cite{Ahma1,Ahma2}, the helicity-dependent part of the LFWF
for vector meson  $K^*$ has been chosen as:
$S_{h,\bar{h}}^{K^{*},\lambda}(\alpha,\mathbf{k})= \frac{\bar
u_{{h}}((1-\alpha)p^+,-\mathbf{k})}{\sqrt{(1-\alpha)}}(\gamma.\varepsilon^*_{\lambda})\,\frac{{v}_{\bar{h}}
(\alpha\,p^+ , \mathbf{k} )}{\sqrt{\alpha}}, $ in analogy with
vector meson, we propose $S_{h,\bar{h}}^{K_{1A},\lambda}$ for the
axial-vector state $K_{1A}$ as:
\begin{eqnarray}\label{eq214}
S_{h,\bar{h}}^{K_{1A},\lambda}(\alpha,\mathbf{k})= \frac{\bar
u_{{h}}((1-\alpha)p^+,-\mathbf{k})}{\sqrt{(1-\alpha)}}(\gamma.\varepsilon^*_{\lambda})\,\gamma_{5}\,\frac{{v}_{\bar{h}}
(\alpha\,p^+ , \mathbf{k} )}{\sqrt{\alpha}}.
\end{eqnarray}
After some calculations and using expressions for $\bar{u}_{h}$ and
$v_{\bar h}$ in light-front coordinate, we extract the factor
$S_{h,\bar{h}}^{K_{1A},\lambda}(\alpha,\mathbf{k})$ as
\begin{eqnarray}\label{eq215}
S_{h,\bar{h}}^{K_{1A}, \pm}(\alpha, \mathbf{k})&= &\pm
\frac{\sqrt{2}}{\alpha\,(1-\alpha)} \Bigg\{ [ (1-\alpha)
\delta_{h\mp,\bar{h}\pm} + \alpha\, \delta_{h\pm,\bar{h}\mp}] {k}
\,e^{\pm i \theta_k} - [(1-\alpha)\,m_{u}\,-\,\alpha\,m_{\bar{s}}]
\delta_{h\pm,\bar{h}\pm}\Bigg\},\nonumber\\
S_{h,\bar{h}}^{K_{1A}, L}(\alpha,
\mathbf{k})&=&-\frac{1}{m_{_{K_{1A}}}
\alpha\,(1-\alpha)}\Bigg\{[\,\alpha\,(1-\alpha)\,m_{K_{1A}}^{2}\,+\mathbf{k}^{2}-m_{u}\,m_{\bar{s}}]\,
(\delta_{h{-},\bar{h}{+}}-\delta_{h{+},\bar{h}{-}})\nonumber\\
&+&\left.{k}\,[{m_{u}+m_{\bar s}}]\,(e^{-i \theta_k}\,
\delta_{h{+},\bar{h}{+}}+e^{i
\theta_k}\,\delta_{h{-},\bar{h}{-}})\right.\Bigg\}.
\end{eqnarray}
In this relation, we have  used the polar representation of the
transverse momentum, i.e. $\mathbf{k}=k\,e^{ i \theta_k}$.  Using
Eqs. (\ref{eq211}), (\ref{eq212}) and (\ref{eq215}), we can rewrite
Eqs. (\ref{eq28}) and (\ref{eq29}) as:
\begin{eqnarray}
\Phi_{K_{1A}}^\parallel(\alpha,\mu) &=&\frac{N_c}{\pi f_{_{K_{1A}}}
m_{_{K_{1A}}}} \int^{|\textbf{k}|<\mu} \frac{d^2
\textbf{k}}{(2\pi)^2} [\alpha\,(1-\alpha)\, m_{_{K_{1A}}}^2
-m_{u}\,m_{\bar{s}}
+\textbf{k}^2] \frac{\phi^{K_{1A}}_{L}(\alpha, \textbf{k})}{\alpha(1-\alpha)},\label{eq216}\\
\Phi_{K_{1A}}^{\perp}(\alpha,\mu) &=&\left.\frac{N_c}{2\pi
f_{_{K_{1A}}}^{\perp}} \int^{|\textbf{k}|<\mu}  \frac{d^2
\textbf{k}}{(2\pi)^2} [\,(1-\alpha)\,m_{u}-\,\alpha\,m_{\bar{s}}]
\frac{\phi^{K_{1A}}_{T}(\alpha,
\textbf{k})}{\alpha(1-\alpha)}.\right.\label{eq217}
\end{eqnarray}
Inserting the Fourier transform relations as
\begin{eqnarray*}\label{eq218}
\phi_{\lambda}(\alpha, \textbf{k})=\int d^2 \textbf{r} \, e^{-i
\textbf{k}. \textbf{r}}\, \phi_{\lambda}(\alpha, \textbf{r}),~~~~
\textbf{k}^2 \phi_{\lambda}(\alpha, \textbf{k})=\int d^2 \textbf{r}
\, e^{-i \textbf{k}. \textbf{r}}\,(-\nabla^{2})
\phi_{\lambda}(\alpha, \textbf{r}) ,
\end{eqnarray*}
into Eqs. (\ref{eq216}) and (\ref{eq217}) and using relations such
as $\int_{0}^{2\pi} e^{-i {k}{r} cos\theta }d\theta =2\pi J_{0}(k
r)$, and $\int_{0}^{\mu} k\, J_{0}(kr)\,dk ={\mu}/{r}\,J_{1}(\mu
r)$, where $J_{0}$ and $J_{1}$ are Bessel functions, we obtain the
following expressions for the twist-2 DAs of $K_{1A}$ state  as:
\begin{eqnarray}
\Phi_{K_{1A}}^\parallel(\alpha,\mu) &=&\frac{N_c}{\pi f_{_{K_{1A}}}
m_{_{K_{1A}}}} \int dr \; \mu J_1(\mu r)
[\alpha\,(1-\alpha)\,m_{_{K_{1A}}}^2  -m_{u}\,m_{\bar{s}}
-\nabla_r^2]
\frac{\phi^{K_{1A}}_{L}(r,\alpha)}{\alpha(1-\alpha)},\label{eq219}\\
\Phi_{K_{1A}}^{\perp}(\alpha,\mu) &=&\left.\frac{N_c}{\pi
f_{_{K_{1A}}}^{\perp}} \int dr \; \mu J_1(\mu r)
[\,(1-\alpha)\,m_{u}-\,\alpha\,m_{\bar{s}}]
\frac{\phi^{K_{1A}}_{T}(r,\alpha)}{\alpha(1-\alpha)}.\right.\label{eq220}
\end{eqnarray}
Similarly, we  can estimate the twist-2 DAs for $K_{1B}$ state as
\begin{eqnarray}
\Phi_{K_{1B}}^\parallel(\alpha,\mu) &=&\frac{N_c}{\pi
f_{_{K_{1B}}}^{\perp}} \int dr \; \mu
J_1(\mu r) [\,(1-\alpha)\,m_{u}-\,\alpha\,m_{\bar{s}}] \frac{\phi^{K_{1B}}_{L}(r, \alpha)}{\alpha(1-\alpha)},\label{eq221}\\
\Phi_{K_{1B}}^{\perp}(\alpha,\mu) &=&\left.\frac{N_c}{\pi
f_{_{K_{1B}}} m_{_{K_{1B}}}} \int dr \; \mu J_1(\mu r)
[\alpha\,(1-\alpha)\,m_{_{K_{1B}}}^2  -m_{u}\,m_{\bar{s}}
-\nabla_r^2] \frac{\phi^{K_{1A}}_{T}(r,
\alpha)}{\alpha(1-\alpha)}.\right.\label{eq222}
\end{eqnarray}
Having the twist-2 DAs, we can obtain the twist-3 DAs
$g_{\bot}^{(a)}$, $g_{\bot}^{(v)}$, $h_{\|}^{(t)}$ and
$h_{\|}^{(p)}$  by Wandzura-Wilczek-–type relations as
\cite{Wandzura}
\begin{eqnarray}\label{eq223}
g_{\perp}^{(a)}(u)&\simeq&{1\over 2}\left[ \int_0^udv\,
\frac{{\Phi}^\parallel(v)}{\bar{v}}+ \int_u^1dv\,\frac{{\Phi}^\parallel(v)}{v} \right],\nonumber\\
g_{\perp}^{(v)}(u)&\simeq&2\left[\bar{u} \int_0^udv\,
\frac{{\Phi}^\parallel(v)}{\bar{v}} + u\int_u^1dv\,\frac{{\Phi}^\parallel(v)}{v}\right],\nonumber\\
h_{\parallel}^{(t)}(u) &=& \xi \left[ \int_{0}^{u} dv
\frac{{\Phi}^{\perp}(v)}{\bar v} - \int_{u}^{1} dv
\frac{{\Phi}^{\perp}(v)}{v} \right],\nonumber\\
h_{\parallel}^{(p)}(u) &=& 2 \left[ \bar u \int_{0}^{u} dv
\frac{{\Phi}^{\perp}(v)}{\bar v} + u \int_{u}^{1} dv
\frac{{\Phi}^{\perp}(v)}{v} \right],
\end{eqnarray}
where $\xi = 2u-1$ and $\bar u=1-u$.

Now, we are also able to calculate the  decay constants in terms of
the LFWFs. The G-parity conserving decay constants of the axial
vector-states are defined as:
\begin{eqnarray}
\langle 0|\bar u(0)\, \gamma^\mu \gamma^5\, s(0)| K_{1A}(p, \lambda
)\rangle &=&- i f_{K_{1A}}
m_{K_{1A}}\varepsilon_\lambda^{\mu},\label{eq224}
\\
\langle 0|\bar u(0) \sigma^{\mu\nu}\,\gamma^5\, s(0)|K_{1B}
(p,\lambda)\rangle &=& - f_{K_{1B}}^{\perp}
(\varepsilon^{\mu}_{\lambda} p^{\nu} - \varepsilon^{\nu}_{\lambda}
p^{\mu})\,,\label{eq225}
\end{eqnarray}
and we take $f^{\perp}_{K_{1A}}= f_{K_{1A}}$,
$f_{K_{1B}}=f^{\perp}_{K_{1B}}$ in $\mu=1$ GeV \cite{Yang1,Y1}.
After expanding the left-hand-sides of Eqs. (\ref{eq224}) and
(\ref{eq225} ) the same way as before, we obtain the decay constants
as follows:
\begin{eqnarray}
f_{K_{1A}} &=& \frac{N_c}{m_{K_{1A}} \pi}  \int_0^1 d\alpha
[\alpha\,(1-\alpha)\,m^{2}_{K_{1A}} -m_{u}\,m_{\bar{s}}
-\nabla_{r}^{2}]
\frac{\phi^{K_{1A}}_{L}(r,\alpha)}{\alpha\,(1-\alpha)}
\bigg{|}_{r=0}\,,\label{eq226}\\
f_{K_{1B}}^{\perp} &=& \frac{N_c}{m_{K_{1B}} \pi}  \int_0^1 d\alpha
[\alpha\,(1-\alpha)\,m^{2}_{K_{1B}} -m_{u}\,m_{\bar{s}}
-\nabla_{r}^{2}]
\frac{\phi^{K_{1B}}_{L}(r,\alpha)}{\alpha\,(1-\alpha)}
\bigg|_{r=0}.\label{eq227}
\end{eqnarray}

To specify $\phi^{K_{1A}(K_{1B})}_{\lambda} (r, \alpha)$ which
includes dynamical properties of $K_{1A}$ (or $K_{1B}$) in the LFWF
in Eq. (\ref{eq213}), we are going to use the AdS/QCD. Based on a
first semiclassical approximation to the light-front QCD, with
massless quarks, the function $\phi_{\lambda}$ can be factorized as
\cite{GFdeSJBr}
\begin{eqnarray}\label{eq228a}
\phi_{\lambda}(\zeta,\alpha, \theta) =\mathcal{N}_{\lambda}\,
\frac{\psi(\zeta)}{\sqrt{2 \pi \zeta}}\, f(\alpha) \, e^{i L
\theta},
\end{eqnarray}
where $\mathcal{N}_{\lambda}$ is a normalization constant which
depends on polarization of the axial-vector meson. In this relation,
$L$ is the orbital angular momentum quantum number and variable
$\zeta=\sqrt{\alpha(1-\alpha)}\,r$, where $r$ is the transverse
distance between the quark and anti-quark forming the meson. The
function $\psi(\zeta)$ satisfies the so-called holographic
light-front Schroedinger equation as
\begin{eqnarray}\label{eq228}
\left(-\frac{d^2}{d\zeta^2}-\frac{1-4L^2}{4\zeta^2}+U(\zeta)\right)\psi(\zeta)=M^2
\psi(\zeta),
\end{eqnarray}
where $M$ is hadron bound-state mass and $U(\zeta)$ is the effective
potential which involves all the complexity of the interaction terms
in the QCD Lagrangian.

According the AdS/QCD, the holographic light-front Schroedinger
equation maps onto the wave equation for strings propagating in AdS
space if $\zeta$ is identified with the fifth dimension in
AdS$_{5}$. To illustrate this issue, we start with the generalized
Proca action in AdS$_{5}$ as  \cite{Erlich}
\begin{eqnarray}\label{eq229}
S =   \int \! d^4 x \, dz  \,\sqrt{g} \,e^{\varphi(z)} \left(
\frac{1}{4} g^{M R} g^{N S} F_{M N} F_{R S} -\frac{1}{2} \, \mu^2
g^{M N} \Phi_M \Phi_N \right),
\end{eqnarray}
where $g = {(\frac{R}{z})}^{10}$ is the modulus of the determinant
of the metric tensor $g_{MN}$. The mass $\mu$ in Eq. (\ref{eq229})
is not a physical observable. $\Phi_M(x, z)$ is a vector field and
$F_{MN} =
\partial_M \Phi_N -\partial_N \Phi_M$. In this action, the
dilaton background $\varphi(z)$ is only a function of the
holographic variable $z$ which vanishes if $z \to \infty$. Variation
of Eq. (\ref{eq229}) leads to the system of coupled differential
equations of motion as
\begin{eqnarray}
\left[ \partial_\mu \partial^\mu  - \frac{z^{3}}{e^{\varphi(z)}}
\partial_z \left(\frac{e^{\varphi(z)}}{z^{3}} \partial_z\right)
-\partial_z^2 \varphi + \frac{(\mu R)^2}{z^2} - 3\right] \Phi_z &\! = \!& 0 ,\label{eq230a} \\
\left[ \partial_\mu \partial^\mu - \frac{z}{e^{\varphi(z)}}
\partial_z \left(\frac{e^{\varphi(z)}}{z} \partial_z\right) +
\frac{(\mu R)^2}{z^2}\right] \Phi_\nu & \! =  \! & - \frac{2}{z}
\partial_\nu \Phi_z .\label{eq230b}
\end{eqnarray}
Imposing  the condition $\Phi_z=0$ which means physical hadrons have
no polarization in the $z$ direction, the wave equation is obtained
as
\begin{eqnarray}\label{eq231}
\left[\partial_\mu \partial^\mu-\frac{ z}{e^{\varphi(z)}}
\partial_z \left(\frac{e^{\varphi(z)}}{z} \partial_z\right)
+ \left(\frac{\mu R}{z}\right)^2\right] \Phi_\nu  =  0.
\end{eqnarray}
A free spin-$1$ hadronic state in holographic QCD is described by a
plane wave in physical space-time with polarization components   $
\epsilon_\nu({p})$ along the  physical coordinates and a
$z$-dependent profile function $\Phi_\nu(x, z) = e ^{ i p \cdot x}
\, \Phi(z) \epsilon_\nu({p})$, with invariant mass $p_\mu p^\mu =
M^2$. Inserting $\Phi_\nu(x, z)$ into the wave equation, the
bound-state eigenvalue equation is derived for spin-$1$  hadronic
bound-state as
\begin{eqnarray}\label{eq232}
\left[-\frac{ z}{e^{\varphi(z)}}
\partial_z \left(\frac{e^{\varphi(z)}}{z} \partial_z\right)
+ \left(\frac{\mu R}{z}\right)^2\right] \Phi(z)  =  M^2 \Phi(z).
\end{eqnarray}
Factoring out the scale $\sqrt{z}$ and dilaton factors from the AdS
field as $\Phi=\sqrt{\frac{z}{R}}\,e^{-\varphi(z)/2}\,\psi(z)$, and
using the substitutes $z \to \zeta$, we find  light-front
Schroedinger equation (Eq. (\ref{eq228})) with effective potential $
U(\zeta) = \frac{1}{2} \varphi''(\zeta) +\frac{1}{4}
\varphi'(\zeta)^2  - \frac{1}{\zeta} \varphi'(\zeta)$, and the AdS
mass  $(\mu R)^2 = L^2-1$. In this correspondence, $ \varphi(\zeta)$
and $(\mu R)^2$ are related to the effective potential and the
internal orbital angular momentum $L$, respectively.

Choosing $\varphi(\zeta)=\kappa^2 \zeta^2$ in the soft-wall model
\cite{Karch} leads to $ U(\zeta)=\kappa^4 \zeta^2$. Solving Eq.
(\ref{eq228}) with this potential and comparing the equation for the
quantum mechanical oscillator in polar coordinates, we obtain the
results in eigenfunctions and  eigenvalues as
$\psi(\zeta)=\kappa\,\sqrt{2\zeta}\,e^{-\frac{\kappa^2 \zeta^2}{2}}$
and $ M^2=4\,\kappa^2\,(n+\frac{1+L}{2})$, respectively.

To determine the function $f(\alpha)$ in Eq. (\ref{eq228a}), we use
the condition $ \int_{0}^{1} d\alpha
\frac{f(\alpha)^2}{\alpha\,(1-\alpha)}=1 $ \cite{GFdeSJBr}.
Therefore, $\phi_{\lambda} (r, \alpha)$ for $K_{1A}$ state with
massless quarks, and $n=0$,  $L = 0 $ is obtained as
\begin{eqnarray}\label{eq233}
\phi^{K_{1A}}_{\lambda}(\alpha,\zeta)= \mathcal{N}_{\lambda}\,
\frac{\kappa}{\sqrt{\pi}}\sqrt{\alpha(1-\alpha)} \exp
\left(-\frac{\kappa^2 \zeta^2}{2}\right),
\end{eqnarray}
where $\kappa=M_{K_{1A}}/\sqrt{2}$. To include the light quark
masses, we apply a Fourier transform to $\textbf{k}$-space as
$\widetilde{\phi}(\alpha, \textbf{k}_{\bot})=\int\,
d^2\textbf{r}\,e^{-i\,\textbf{k}_{\bot}.\textbf{r}}\,\phi(\alpha,\zeta)$,
and obtain
\begin{eqnarray}\label{eq235}
{\widetilde{\phi}}^{K_{1A}}_{\lambda}(\alpha, \textbf{k}_{\bot})=
\mathcal{N}_{\lambda}\, \frac{2}{
\sqrt{\alpha(1-\alpha)\,}}\,\frac{\sqrt{\pi}}{\kappa}
\exp\left(-\frac{\textbf{k}_{\bot}^2}{2\alpha(1-\alpha)\,\kappa^2}\right).
\end{eqnarray}
For massive quarks, we should replace \cite{BroTera5}:
\begin{eqnarray}\label{eq236}
\frac{\textbf{k}_{\bot}^2}{\alpha(1-\alpha)}\to
\frac{\textbf{k}_{\bot}^2}{\alpha(1-\alpha)}+
\frac{m_{u}^2}{\alpha}+\frac{m^2_{\bar{s}}}{(1-\alpha)}.
\end{eqnarray}
After substituting this into the wave function and Fourier
transforming back to transverse position-space, one obtains the
final form of the AdS/QCD wave function:
\begin{eqnarray}\label{eq237}
\phi^{K_{1A}}_{\lambda} (\zeta, \alpha) =\mathcal{N}_{\lambda}\,
\frac{\kappa}{\sqrt{\pi}} \, \sqrt{\alpha\,(1-\alpha)} \exp
\left(-\frac{\kappa^2 \zeta^2}{2}\right) \exp\left
\{-\left[\frac{m_u^2-\alpha(m_u^2-m^2_{\bar{s}})}{2\alpha
(1-\alpha)\,\kappa^2} \right] \right \}.
\end{eqnarray}
In position-space, $\mathcal{N}_{\lambda}$ can be fixed by this
normalization condition \cite{R1}:
\begin{equation}\label{eq}
\int d^{2} {\mathbf{r}} \,d\alpha \Bigg[\sum_{h,\bar{h}}
|\Psi^{K_{1A}, \lambda}_{h,\bar{h}}(r, \alpha)|^{2}\Bigg] = 1.
\end{equation}

In the next section, we estimate the decay constants and DAs for
$K_1(1270)$ and $K_1(1400)$ mesons. As an application of these DAs,
we can use them to calculate the transition form factors of the
semileptonic $B\to K_{1}(1270, 1400)\,\ell^{+}\ell^{-}$ decays.

\section{NUMERICAL ANALYSIS}\label{sec.3}
In this section, we present our numerical analysis for the DAs of
$K_1(1270, 1400)$ mesons in terms of the DAs of $K_{1A}$ and
$K_{1B}$ states in the AdS/QCD correspondence. Then, the transition
form factors of $B \to K_1 \ell^{+}\ell^{-}$ decays are
investigated. The other phenomenological quantities can be evaluated
by using these form factors. In this paper, we take masses as:
$m_b=(4.81\pm 0.03)$~GeV, $m_{B}=(5.27\pm0.01)$~GeV \cite{pdg},
$m_{K_{1A}} = (1.31 \pm 0.06)$~GeV, and $m_{K_{1B}} = (1.34 \pm
0.08)$~GeV \cite{Y1}. In addition, we choose light quark masses as
$m_{u,d}=350$~MeV and $m_s=480$~MeV \cite{ Ahma4}. It should be
noted that the values of  the effective quark masses, used in the
holographic LFWFs, are clearly different from the conventional
constituent masses in the non-relativistic theories.

We obtain the decay constant values for $ K_{1A}$ and $ K_{1B} $
states from Eqs. (\ref{eq226}) and (\ref{eq227})  as presented in
Table \ref{T32}. This table also contains the results obtained in
the frame work of the LCSR \cite{Y1}.  As mentioned before, we take
$f^{\perp}_{K_{1A}}= f_{K_{1A}}$, and
$f_{K_{1B}}=f^{\perp}_{K_{1B}}$ in our analysis.
\begin{table}[th]
\caption{ Decay constant values of $ K_{1A}$ and $ K_{1B} $ states
in MeV.}\label{T32}
\begin{ruledtabular}
\begin{tabular}{ccc}
Approach                    & $f_{K_{1A}}$       & $f^{\perp}_{K_{1B}}$ \\
\hline
This work           & ${236} \pm {5}$    & ${220} \pm {5}$          \nonumber     \\
\mbox{LCSR} \cite{Y1}       & ${250} \pm {13}$   & ${190} \pm {10}$
\end{tabular}
\end{ruledtabular}
\end{table}
Using Eq. (\ref{eq23}) and values in Table \ref{T32}, we can
evaluate the decay constant values for mesons $K_1$. In Table
\ref{T33}, we compare our predictions for the decay constants of
$K_1(1270)$ and $K_1(1400)$ mesons with those obtained using the
LCSR approach at $\theta_K = {-(34\pm13)}^{\circ}$. The origin of a
large error in calculation  of the decay constants is due to the
uncertainty in determination of the mixing angle.
\begin{table}[th]
\caption{ Decay constant values of $ K_{1}$ mesons (in MeV) compared
to the LCSR at $\theta_K = {-(34\pm13)}^{\circ}$. }\label{T33}
\begin{ruledtabular}
\begin{tabular}{ccccc}
 Approach & $f_{K_1(1270)}$&$f^{\perp}_{K_1(1270)}$&$f_{K_1(1400)}$&$f^{\perp}_{K_1(1400)}$\\
\hline
This work  &$ -169\pm {39}$     & $ 144\pm {38} $   &  $  157\pm {35}$  &  $172\pm {32}$ \\
LCSR & $ -172\pm {43}$   & $ 117\pm {36}$  &  $  171\pm {35}$   & $159\pm {26}$  \\
\end{tabular}
\end{ruledtabular}
\end{table}

\begin{figure}[th]
\includegraphics[width=6.cm,height=5.95cm]{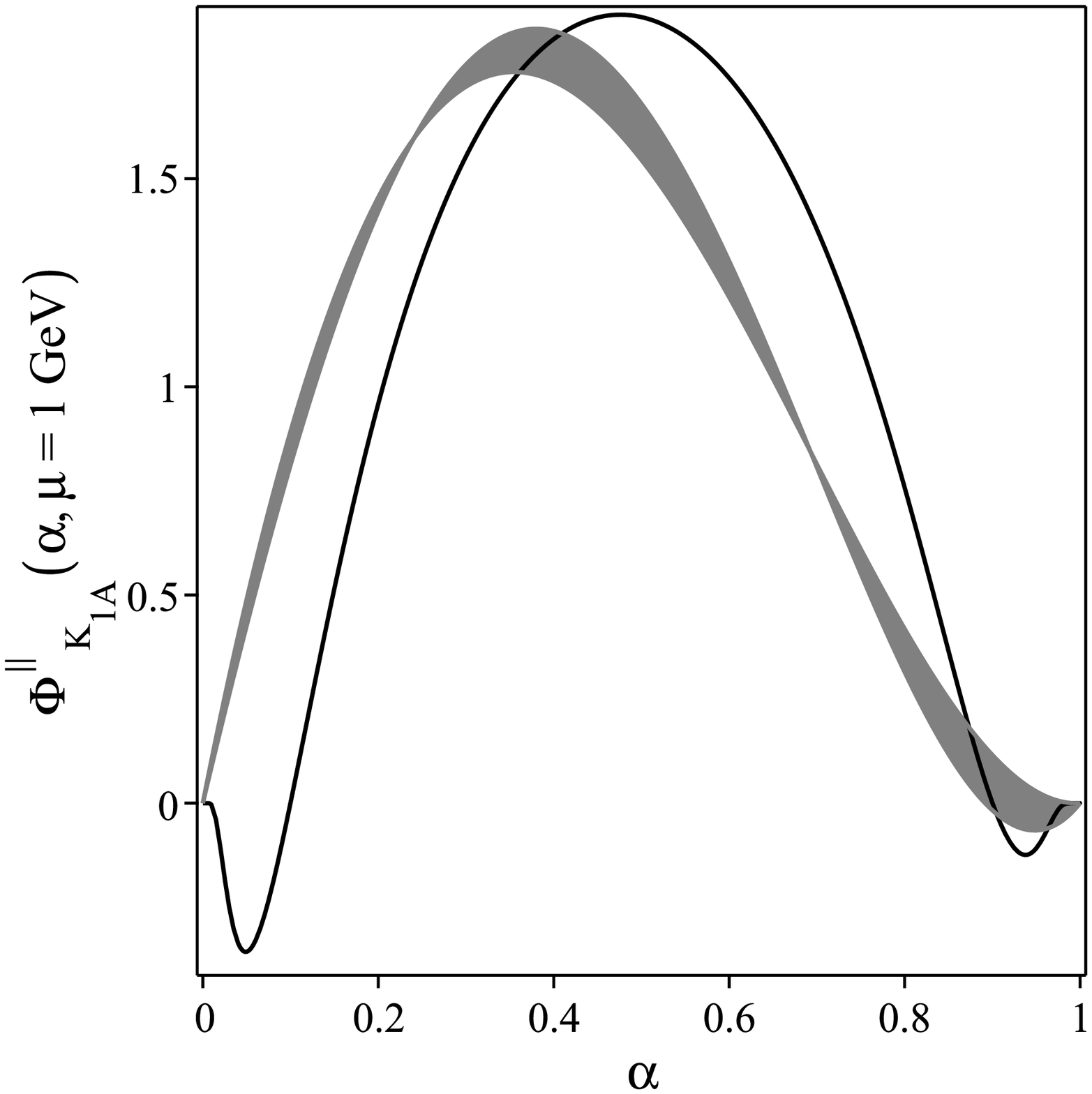}
\includegraphics[width=6.cm,height=5.95cm]{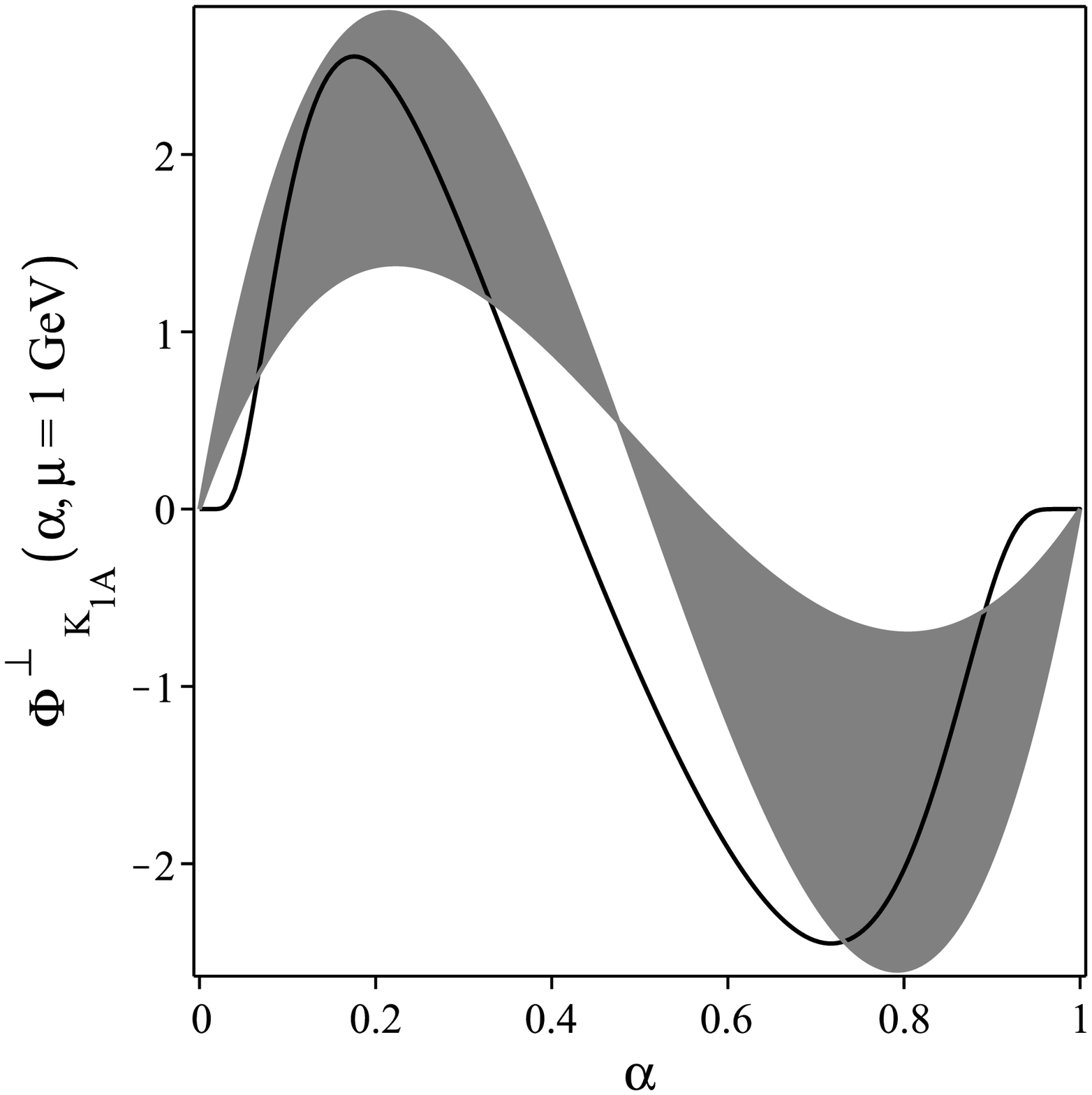}
\includegraphics[width=6.cm,height=5.95cm]{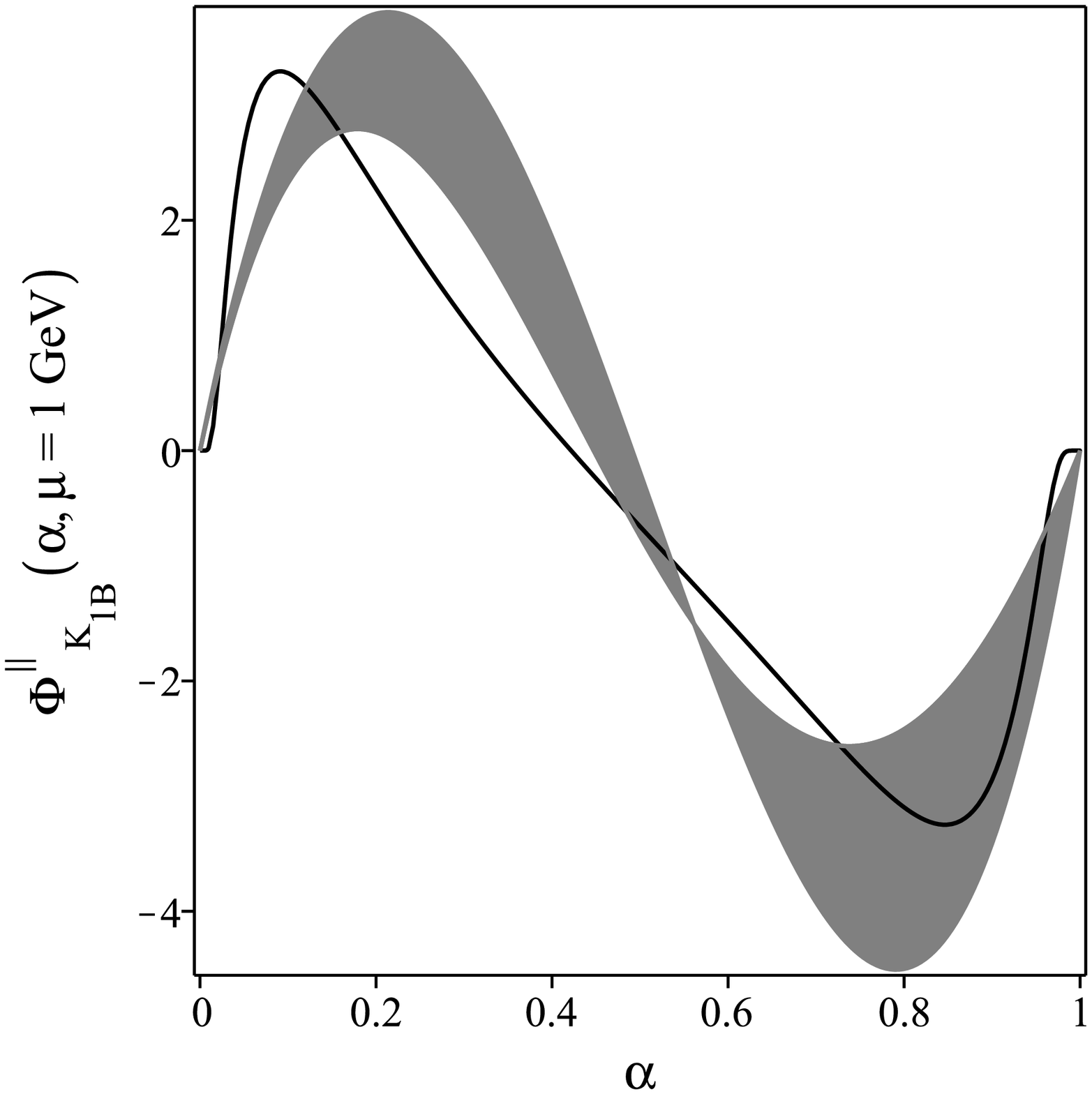}
\includegraphics[width=6.cm,height=5.95cm]{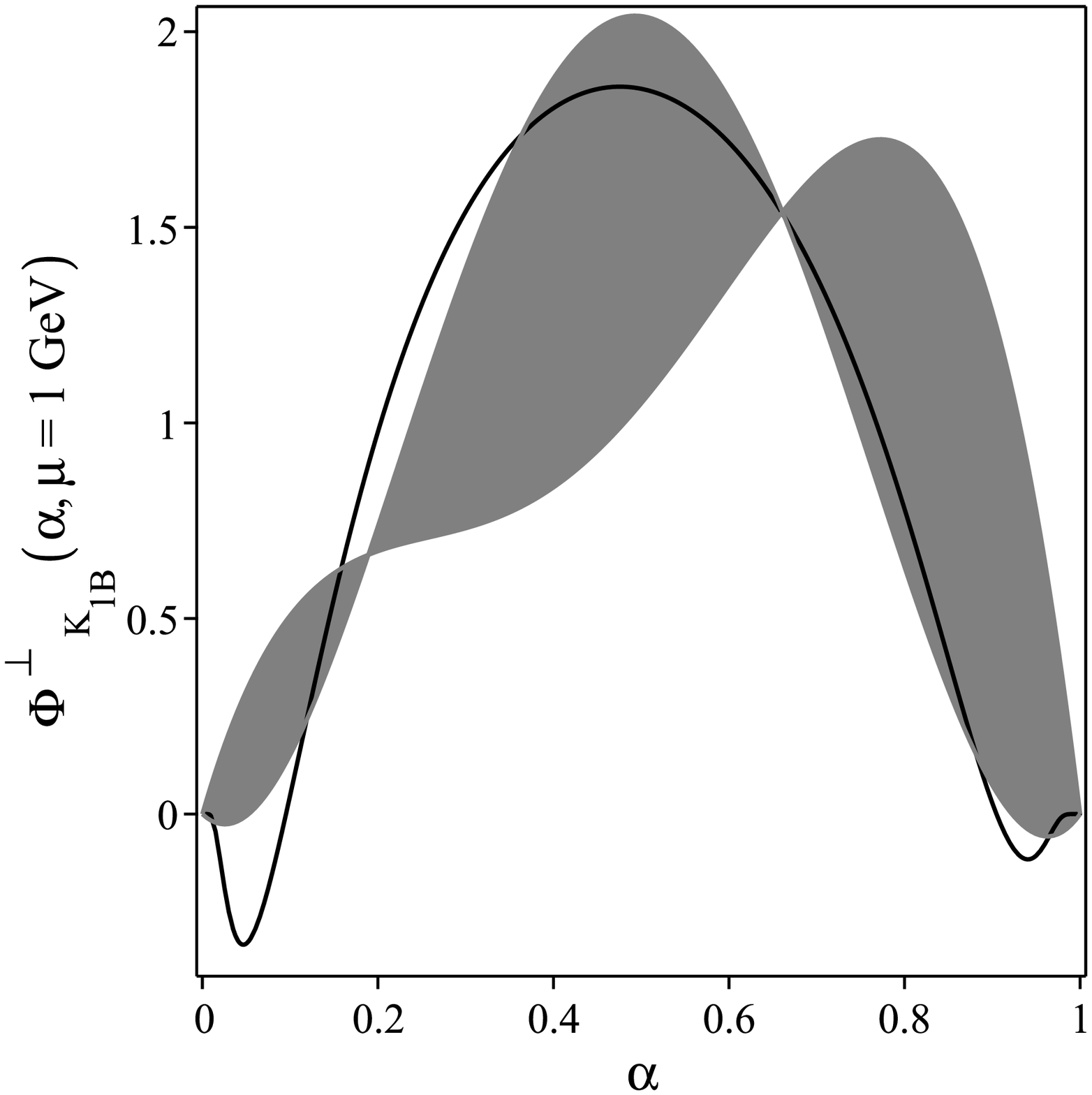}
\caption{The twist-2 DAs for $K_{1A}$ and $K_{1B}$ at $\mu=1 $ ~GeV
in the AdS/QCD. Gray areas show the LCDAs by considering their
errors. }\label{F31}
\end{figure}

\begin{figure}[th]
\includegraphics[width=6.cm,height=5.95cm]{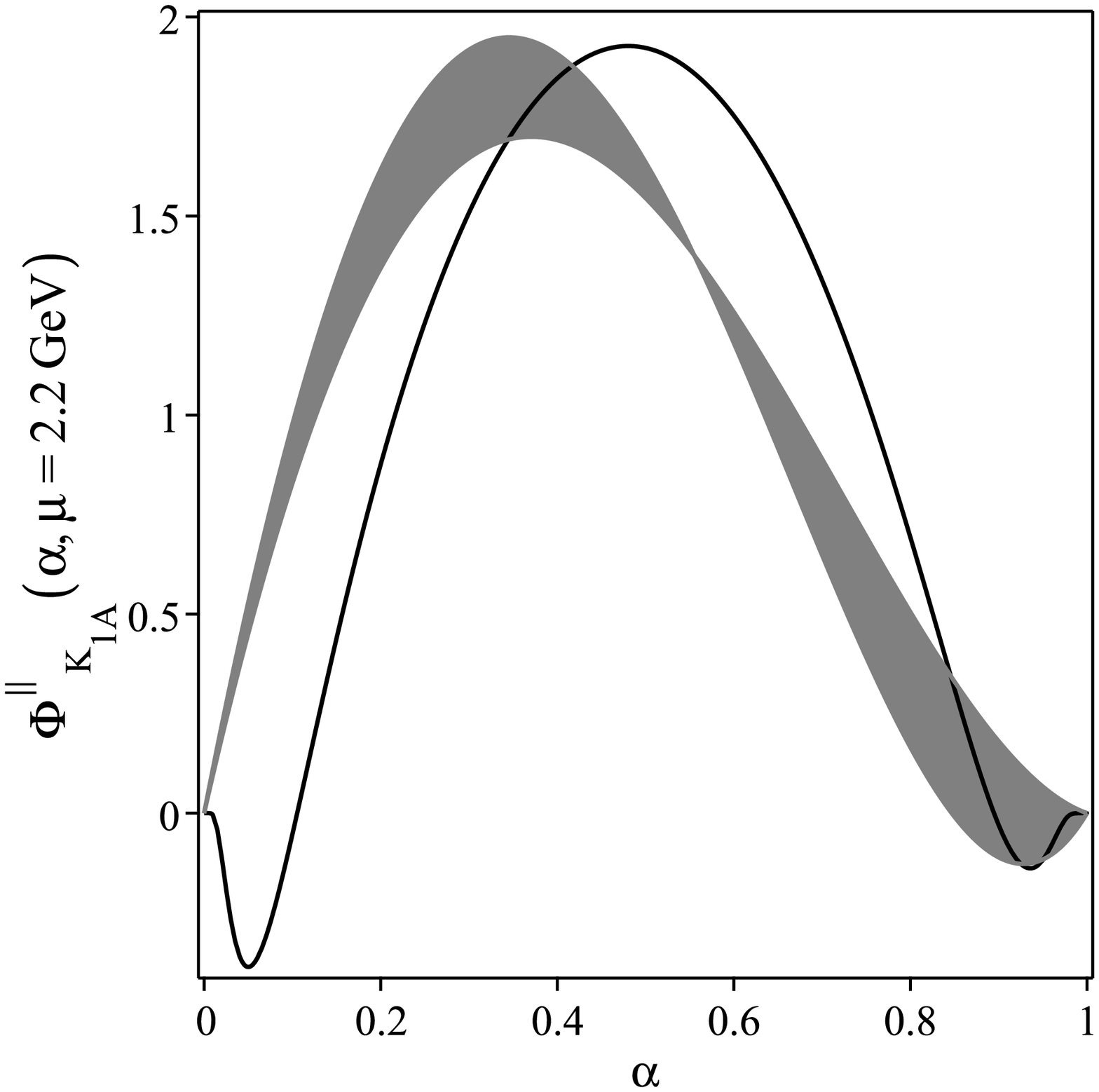}
\includegraphics[width=6.cm,height=5.95cm]{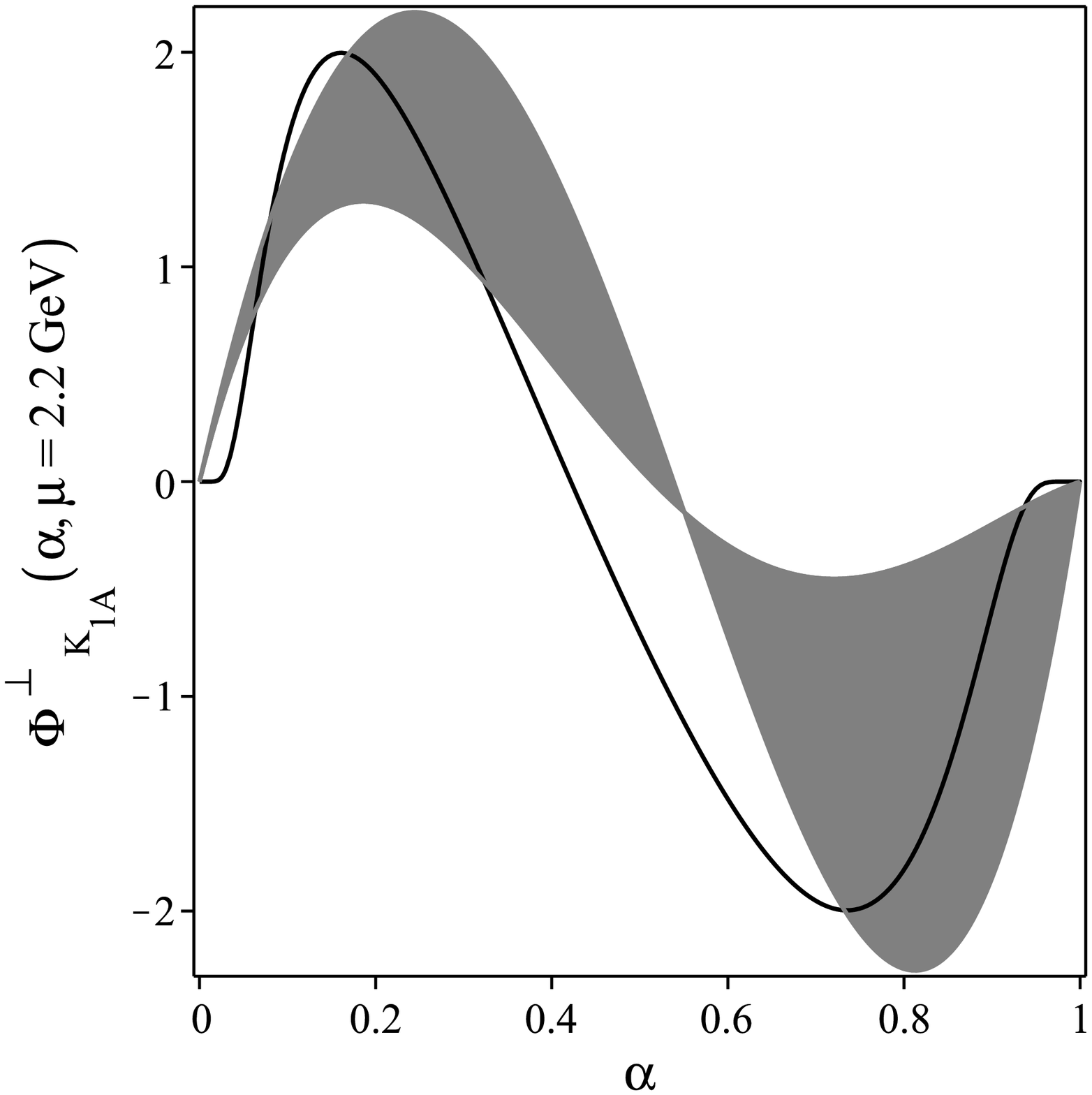}
\includegraphics[width=6.cm,height=5.95cm]{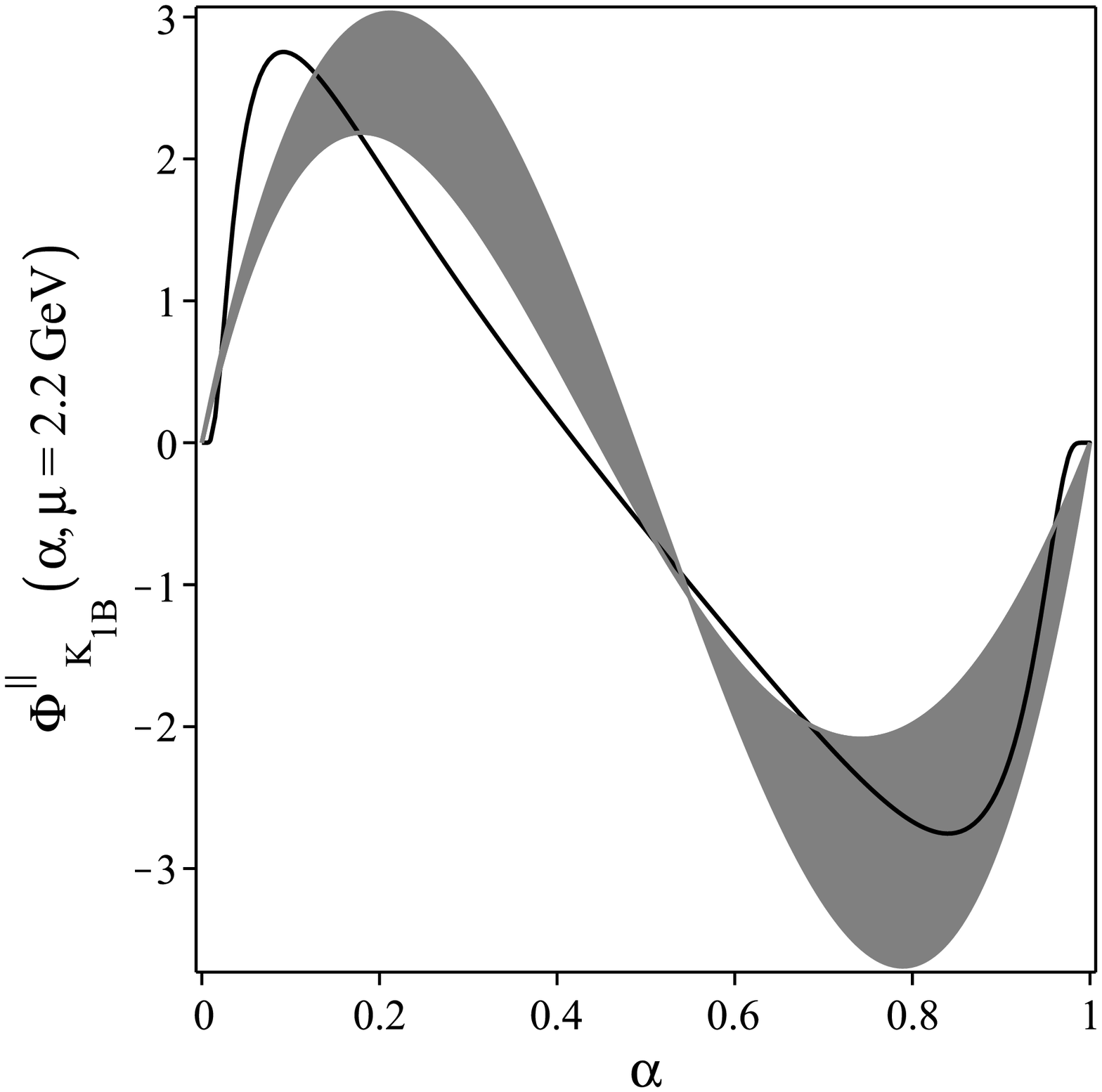}
\includegraphics[width=6.cm,height=5.95cm]{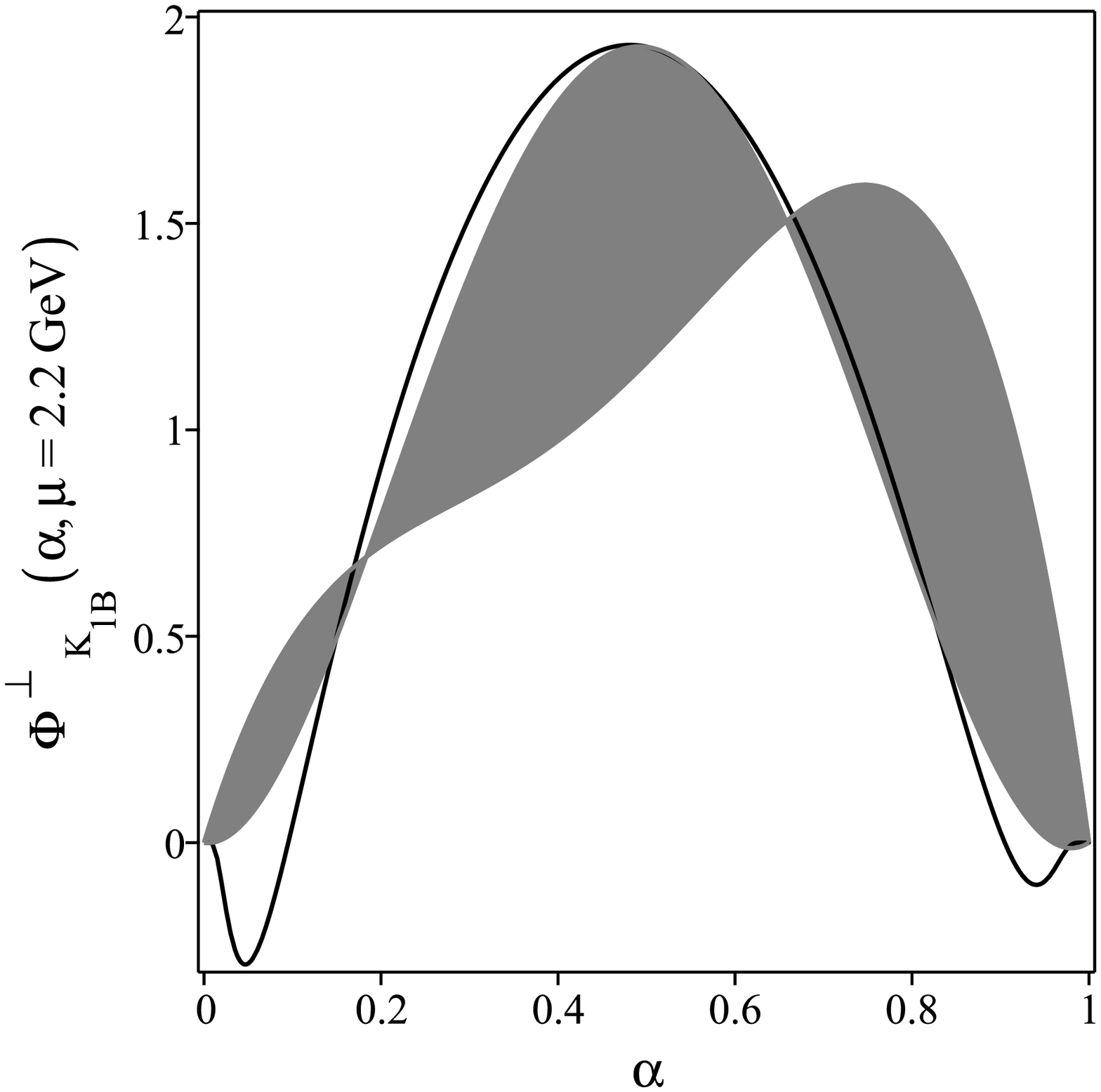}
\caption{The same as Fig. \ref{F31} but for  $\mu=2.2$~GeV.
}\label{F32}
\end{figure}

\begin{figure}[th]
\includegraphics[width=6.cm,height=5.95cm]{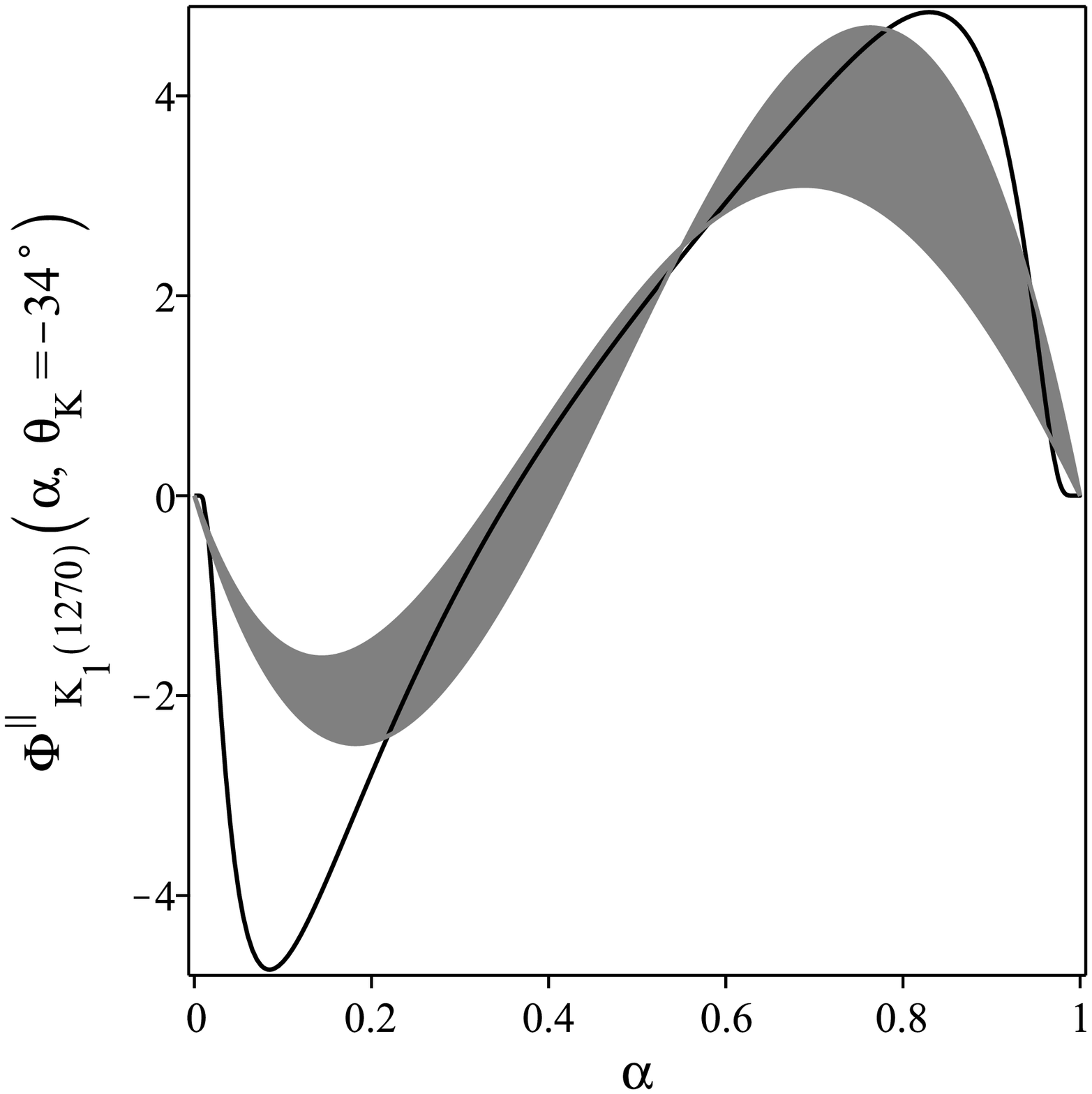}
\includegraphics[width=6.cm,height=5.95cm]{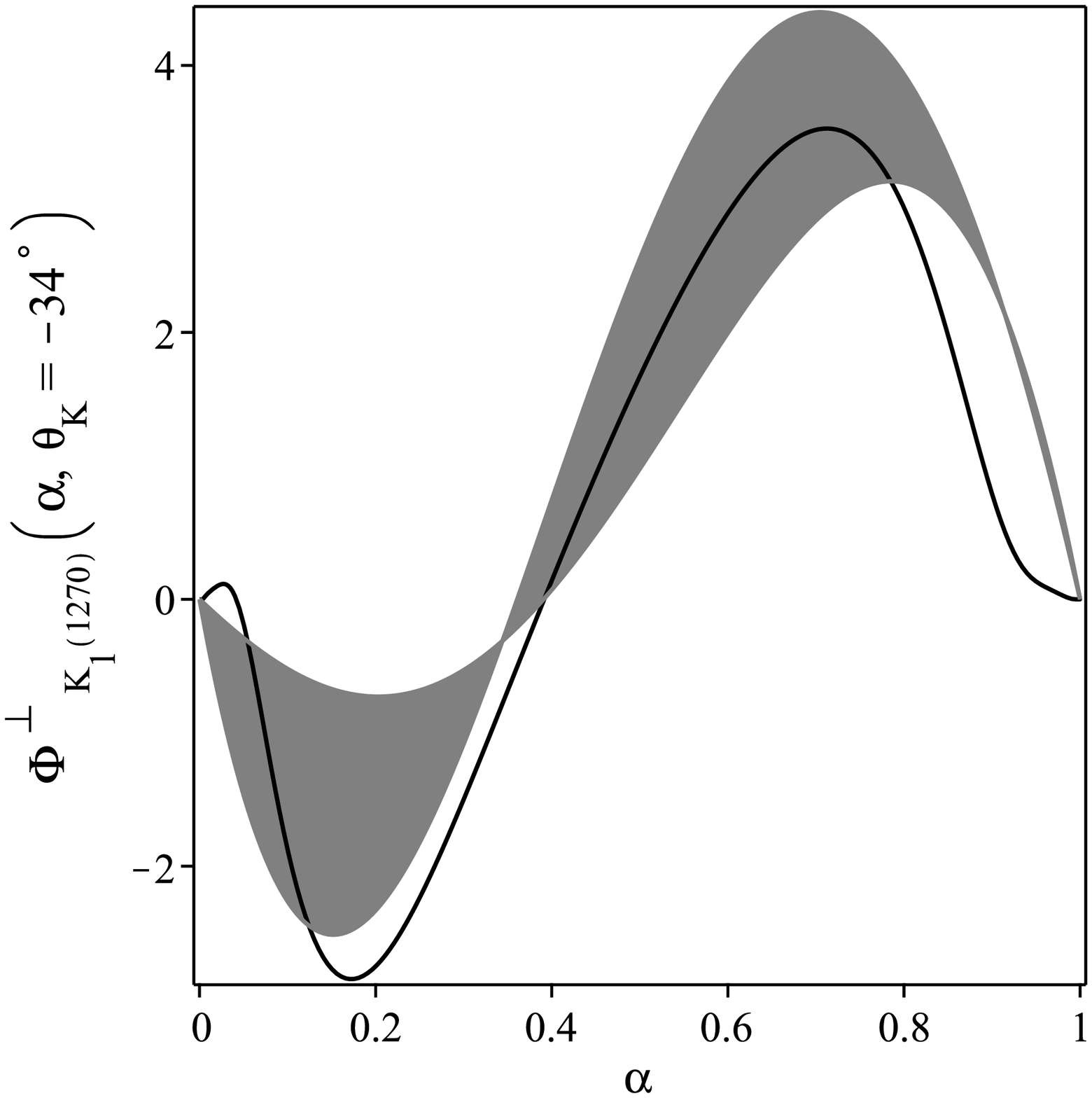}
\includegraphics[width=6.cm,height=5.95cm]{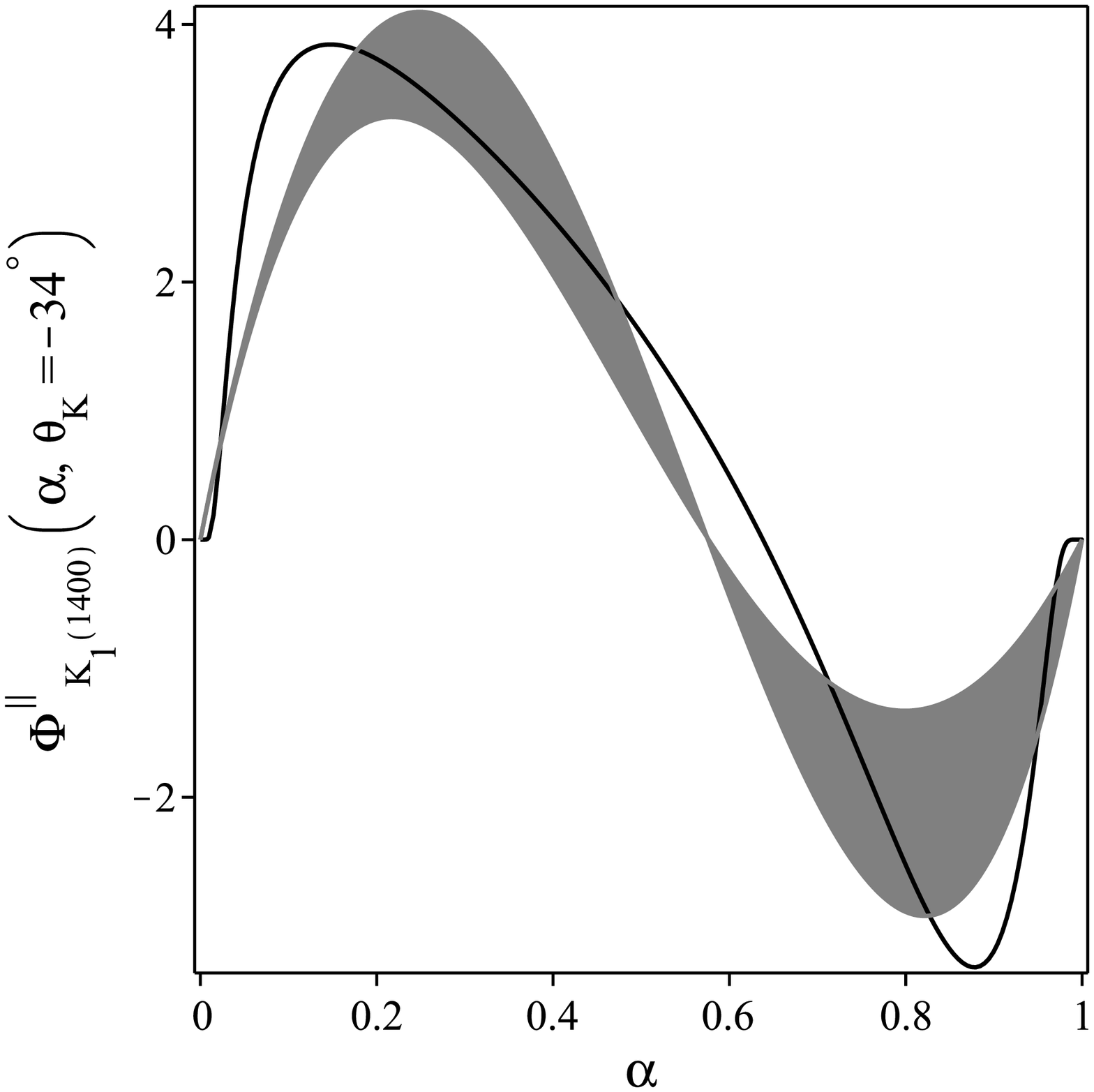}
\includegraphics[width=6.cm,height=5.95cm]{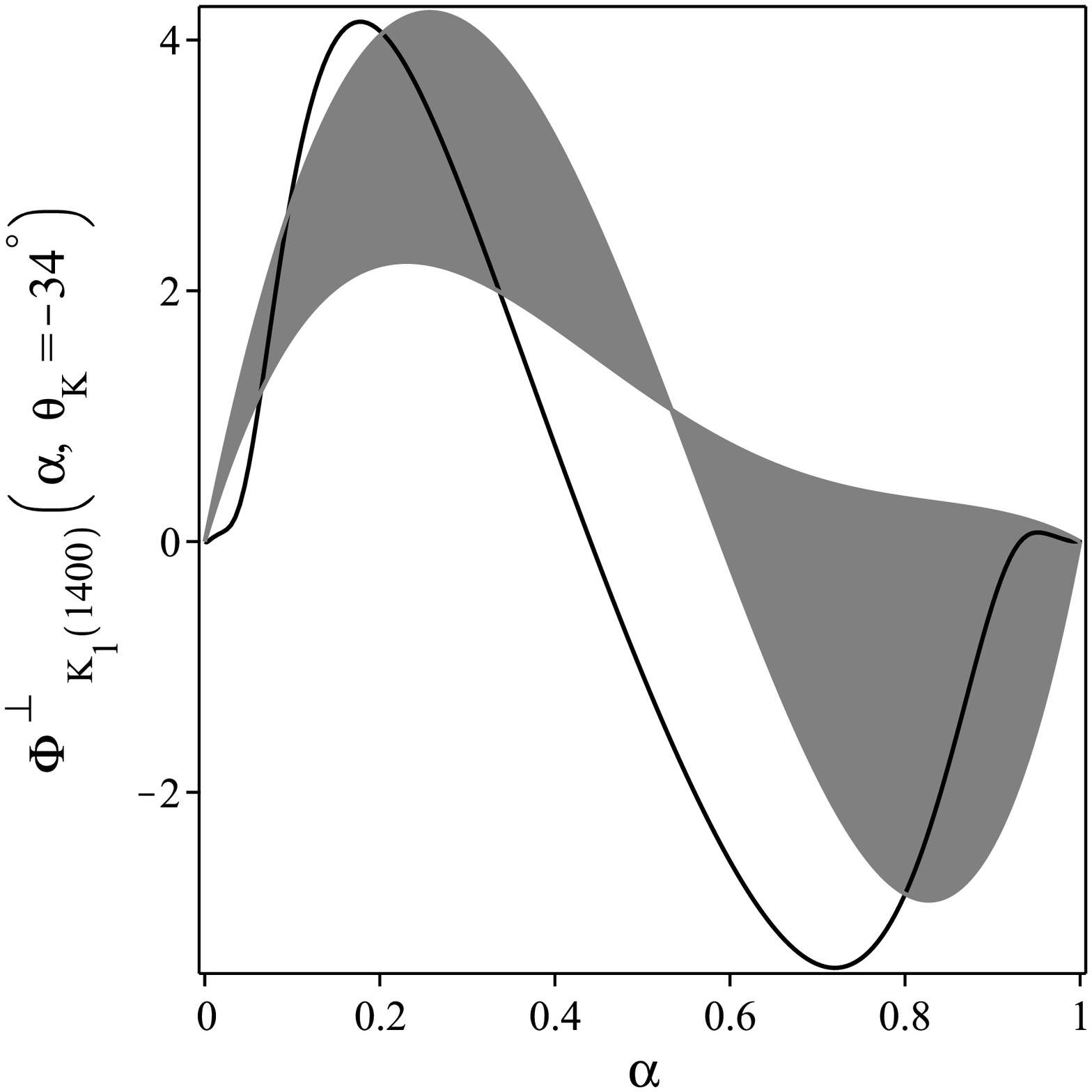}
\caption{The twist-2 DAs for $K_{1}(1270)$ and $K_{1}(1400)$ mesons
at $\mu=1 $ ~GeV and $\theta_K=- 34^{\circ} $ in the AdS/QCD. Gray
areas show the LCDAs by considering their errors. }\label{F33}
\end{figure}

The approximate forms of the twist-2 DAs for $K_{1A}$ and $K_{1B}$
states in the frame work of the LCSR are as follows:
\begin{eqnarray}
\Phi^{\parallel,\perp}(u) & = & 6 u \bar u \left[
a_{0}^{\parallel,\perp}\, + 3 a_{1}^{\parallel,\perp}\, \xi +
a_{2}^{\parallel,\perp}\, \frac{3}{2} ( 5\xi^2  - 1 ) \right],
\end{eqnarray}
where $\xi=2u-1$. The values of the Gegenbauer moments
$a_{i}^{\parallel,\perp}~ i=(0, 1, 2)$, for two states $K_{1A}$ and
$K_{1B}$ have been estimated in Ref. \cite{Yang1} and given in Table
\ref{T34}.
\begin{table}[th]
\caption{Gegenbauer moments of $\Phi_\parallel$ and $\Phi_\perp$ for
$K_{1A}$ and $K_{1B}$ states.} \label{T34}
\begin{ruledtabular}
\begin{tabular}{ccccccc}
$\mu$ &${a_0^{\parallel, K_{1A}}}$&  ${a_1^{\parallel, K_{1A}}}$&
$a_2^{\parallel, K_{1A}}$ & $a_0^{\perp, K_{1A}}$    & $ a_1^{\perp,
K_{1A}}$ &
$a_2^{\perp,K_{1A}}$\\
\hline ${\rm 1~GeV}$  & $1$ & ${-0.30}_{-0.20}^{+0.00}$ & $-0.05\pm
0.03$ & $ {0.27}_{-0.17}^{+0.03}$  & $-1.08\pm 0.48$    &  $ 0.02\pm
0.20$
\\ ${\rm 2.2~GeV}$   &  $1$   & ${-0.25}_{-0.17}^{+0.00}$ & $-0.04 \pm 0.02$ &  ${0.25}_{-0.16}^{+0.03}$ & $-0.88\pm 0.39 $ &     $ 0.01\pm 0.15$\\
\hline &$ a_0^{\parallel, K_{1B}}$ & $ a_1^{\parallel, K_{1B}}$ &
$a_2^{\parallel, K_{1B}}$&${a_0^{\perp, K_{1B}}}$&  ${a_1^{\perp,
K_{1B}}}$&$a_2^{\perp,
 K_{1B}}$
\\
\hline ${\rm 1~GeV}$   & $- 0.19\pm 0.07$   & $ -1.95\pm 0.45$  &
$0.10^{+0.15}_{-0.19}$   &$1$ & $0.30^{+0.00}_{-0.33}$& $ -0.02\pm
0.22$
\\   ${\rm 2.2~GeV}$&$-0.19\pm 0.07$&$-1.57\pm 0.37$&$ 0.07^{+0.11}_{-0.14}$&$1$
&$0.24^{0.00}_{-0.27} $&$ -0.02 \pm 0.17$
\end{tabular}
\end{ruledtabular}
\end{table}
Using Eqs. (\ref{eq218})-(\ref{eq222}), and the decay constant
values presented in Table \ref{T33}, we display our predictions for
the twist-2 holographic LFDAs of $K_{1A}$ and $K_{1B}$ states at the
scale $\mu = 1$ and $\mu = 2.2$~GeV in Figs. \ref{F31} and
\ref{F32}, respectively. In these figures, gray areas show the DAs
predicted from the LCSR method for aforementioned states by
considering their errors.
In addition, we  illustrate in Fig. \ref{F33} the two-parton DAs of
twist-2 for $K_1(1270)$ and $K_1(1400)$ mesons at the scale $ \mu =
1$ ~ GeV in the frame work of the AdS/QCD and LCSR, where
$\theta_K=-34^{\circ} $.

Now, the transition form factors of the semileptonic FCNC decays $B
\to K_1(1270, 1400)$, which have been calculated in the LCSR
approach \cite{MoKh}, are evaluated using the holographic DAs. The
explicit expressions of these transition form factors in terms of
the DAs are given in Appendix. We find that, for $s_{0}\simeq
(33\sim 36)$, all considered form factors in the AdS/QCD exhibit
good stability within the Borel mass parameter
$5\,\mbox{GeV}^{2}\leq \,M^2\,\leq 10\, \mbox{GeV}^{2}$. To evaluate
the form factors in the physical region $4\,m_{\ell}^2\,\leq
q^2\leq\, (m_{B}-m_{K_1})^2$, we fit the double-pole form
\begin{eqnarray}\label{eq041}
F_{k}(q^{2})=\frac{F_{k}(0)}{1-\alpha\,(q^2/m_{B}^2)+\beta\,(q^4/m_{B}^4)},
\end{eqnarray}
for each form factor. In this fit function, we use the notation
$F_{k}(q^2)$ to denote the form factors, $F_{k}(0)$,  $\alpha$ and
$\beta$ are the corresponding coefficients and their values are
presented in Table. \ref{T35}  at $\theta_K=-34^\circ$.
\begin{table}[th]
\caption{$F_{k}(0)$,  $\alpha$ and $\beta$ parameters for $B\to
K_{1}(1270)[ K_{1}(1400)]$ form factors using the holographic DAs
in $\theta_K=-34^\circ$. } \label{T35}
\begin{ruledtabular}
\begin{tabular}{cccccccccc}
$F^{B \to K_{1}(1270)}$&$F(0)$&$\alpha$&$\beta$&$F^{B \to K_{1}(1400)}$&$F(0)$&$\alpha$&$\beta$\\
\hline
$A$      &$-0.60$    &$1.49$    &$0.59$  &$A$   &$0.12$    &$2.01$    &$1.41$   \\
$V_0$    &$0.29$    &$2.14$    &$1.20$ &$V_0$   &$-0.29$    &$2.40$    &$1.58$  \\
$V_1$    &$-0.45$    &$0.84$    &$0.10$  &$V_1$   &$0.13$    &$0.77$    &$1.76$  \\
$V_2$   &$-0.39$    &$0.90$    &$0.55$  &$V_2$   &$0.20$    &$1.93$    &$1.72$    \\
$T_1$   &$-0.37$    &$2.64$    &$1.92$  &$T_1$   &$0.11$    &$1.12$    &$1.01$   \\
$T_2$   &$-0.36$    &$0.94$    &$-0.18$  &$T_2$   &$0.10$    &$2.43$    &$1.89$   \\
$T_3$  &$-0.22$    &$-0.15$    &$-0.99$  &$T_3$   &$0.14$    &$2.17$    &$1.93$ \\
\end{tabular}
\end{ruledtabular}
\end{table}
We compare the AdS/QCD predictions for the transition form factors
at $q^2=0$ with those of the LCSR in Table. \ref{T36}. As can be
seen, there is a logical agreement between the AdS/QCD and LCSR
predictions.
\begin{table}[th]
\caption{Our predictions for the form factors in $q^2=0$ compared to
the LCSR predictions in  $\theta_{K}=-34^{\circ}$.} \label{T36}
\begin{ruledtabular}
\begin{tabular}{ccccccccc}
& $B\to K_{1}(1270)$  &   AdS/QCD &   LCSR  & $B\to K_{1}(1400)$  & AdS/QCD & LCSR   \\
\hline
& $A$         &  $-0.60\pm{0.08}$   &$-0.66\pm0.13$   &$A$          &  $0.11\pm0.02$ &$0.14\pm0.03$        \\
& $V_{0}$     &  $0.29\pm0.03$  &$0.24\pm0.04$  &$V_{0}$      &  $-0.29\pm0.02$  &$-0.22\pm0.04$         \\
& $V_1$       &  $-0.45\pm0.06$   &$-0.47\pm0.08$   &$V_1$        &  $0.13\pm0.02$ &$0.18\pm0.03$        \\
& $V_2$       &  $-0.39\pm0.04$   &$-0.39\pm0.06$   &$V_2$        &  $0.20\pm0.02$ &$0.30\pm0.05$        \\
&$T_1$   &  $-0.37\pm0.03$   &$-0.41\pm0.05$   &$T_1$   &  $0.11\pm0.02$ &$0.10\pm0.02$        \\
&$T_2$   &  $-0.36\pm0.03$   &$-0.40\pm0.05$   &$T_{2}$   &  $0.10\pm0.02$ &$0.11\pm0.02$        \\
& $T_3$       &  $-0.22\pm0.02$   &$-0.26\pm0.04$   &$T_3$        &  $0.14\pm0.03$ & $0.16\pm0.04$       \\
\end{tabular}
\end{ruledtabular}
\end{table}

\begin{figure}[th]
\includegraphics[width=7cm,height=7cm]{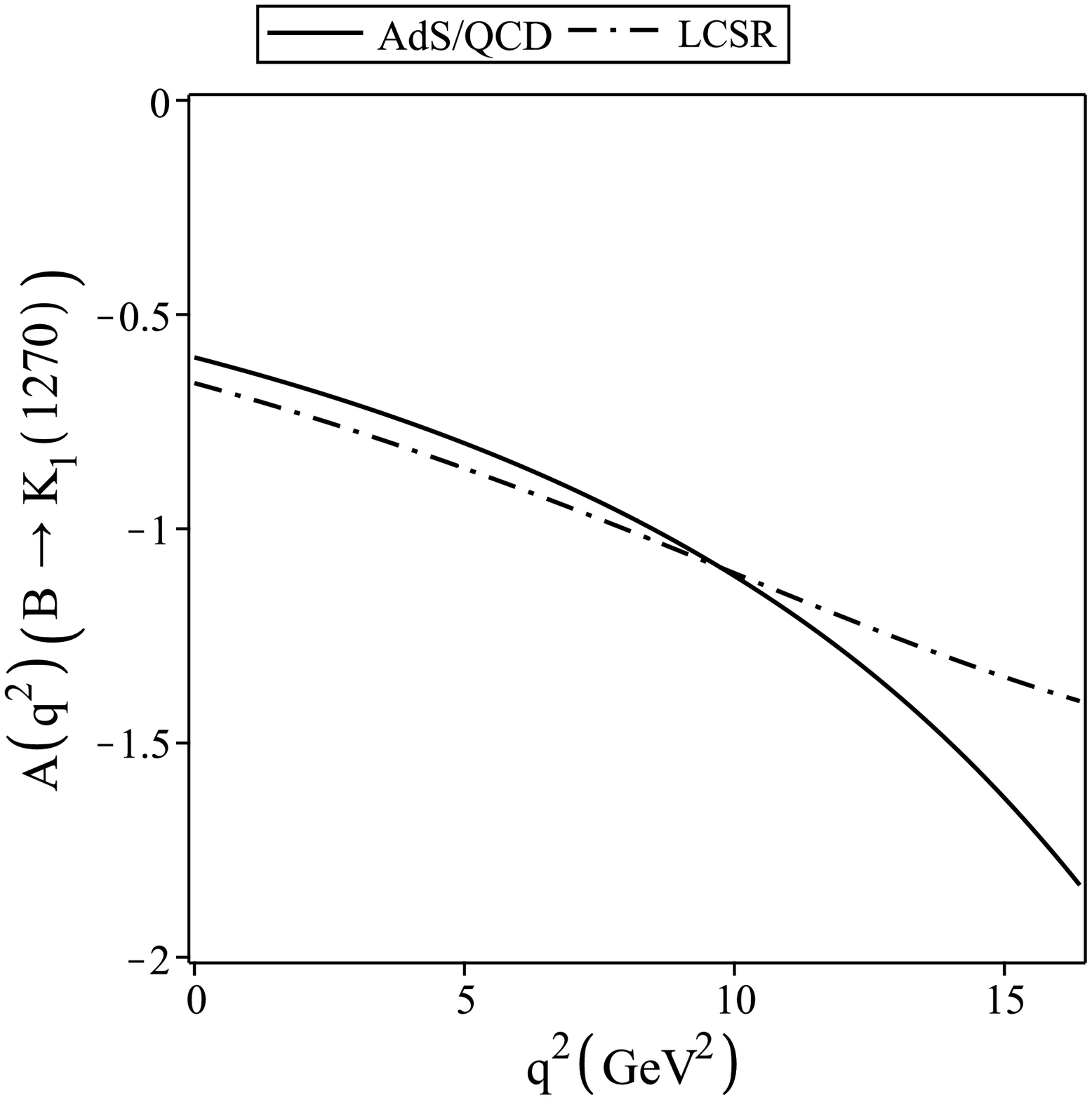}
\includegraphics[width=7cm,height=7cm]{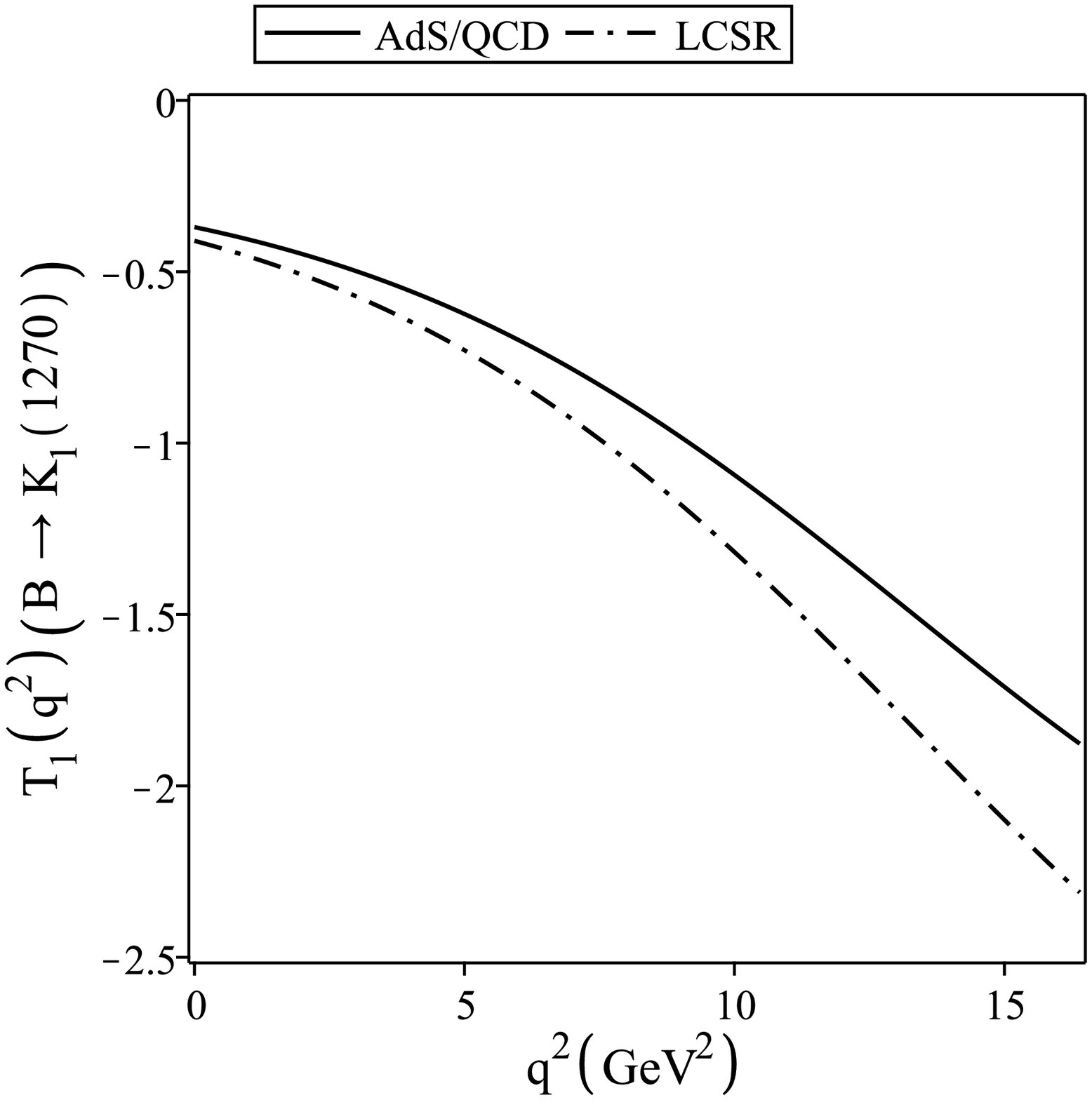}
\includegraphics[width=7cm,height=7cm]{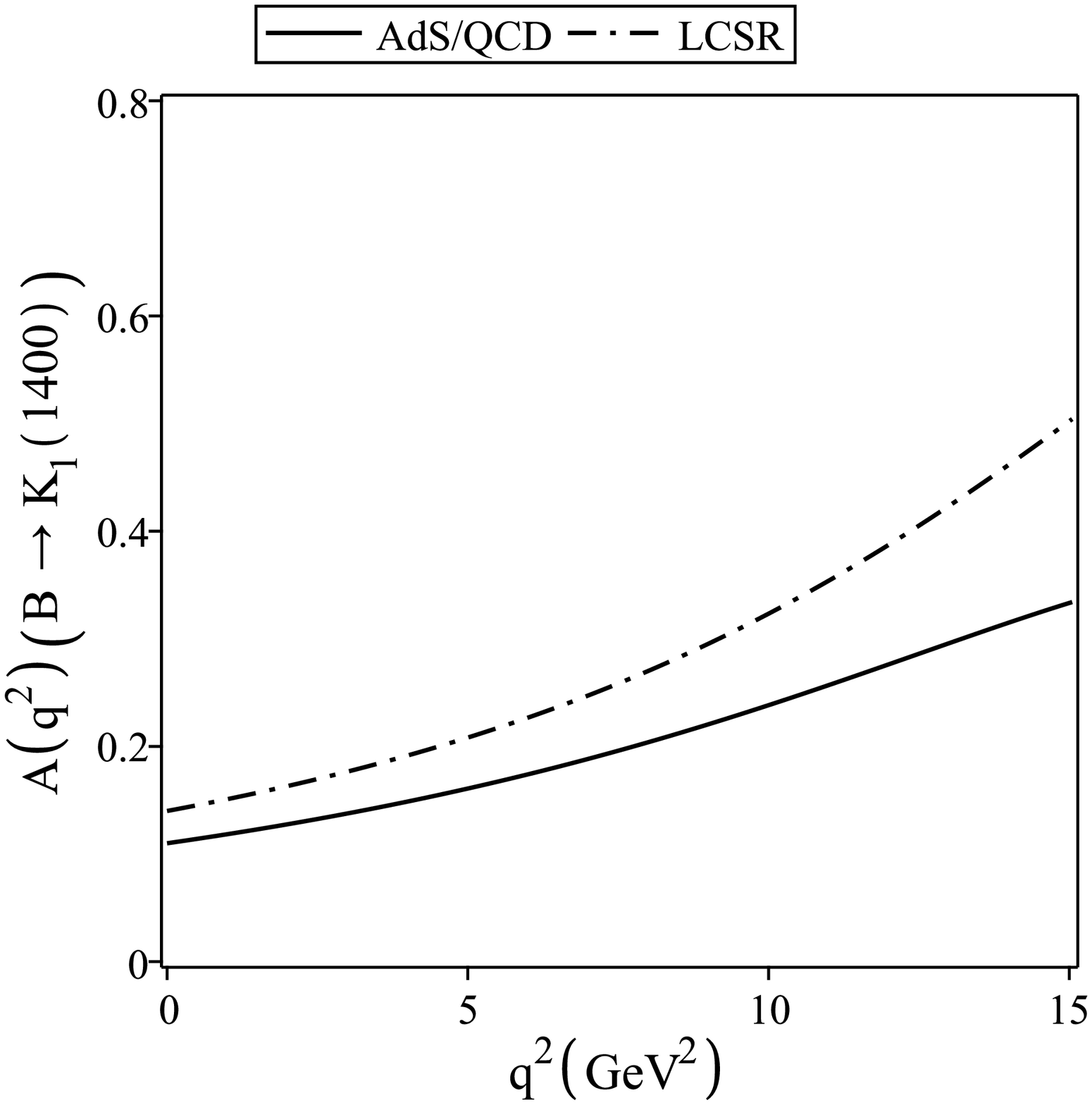}
\includegraphics[width=7cm,height=7cm]{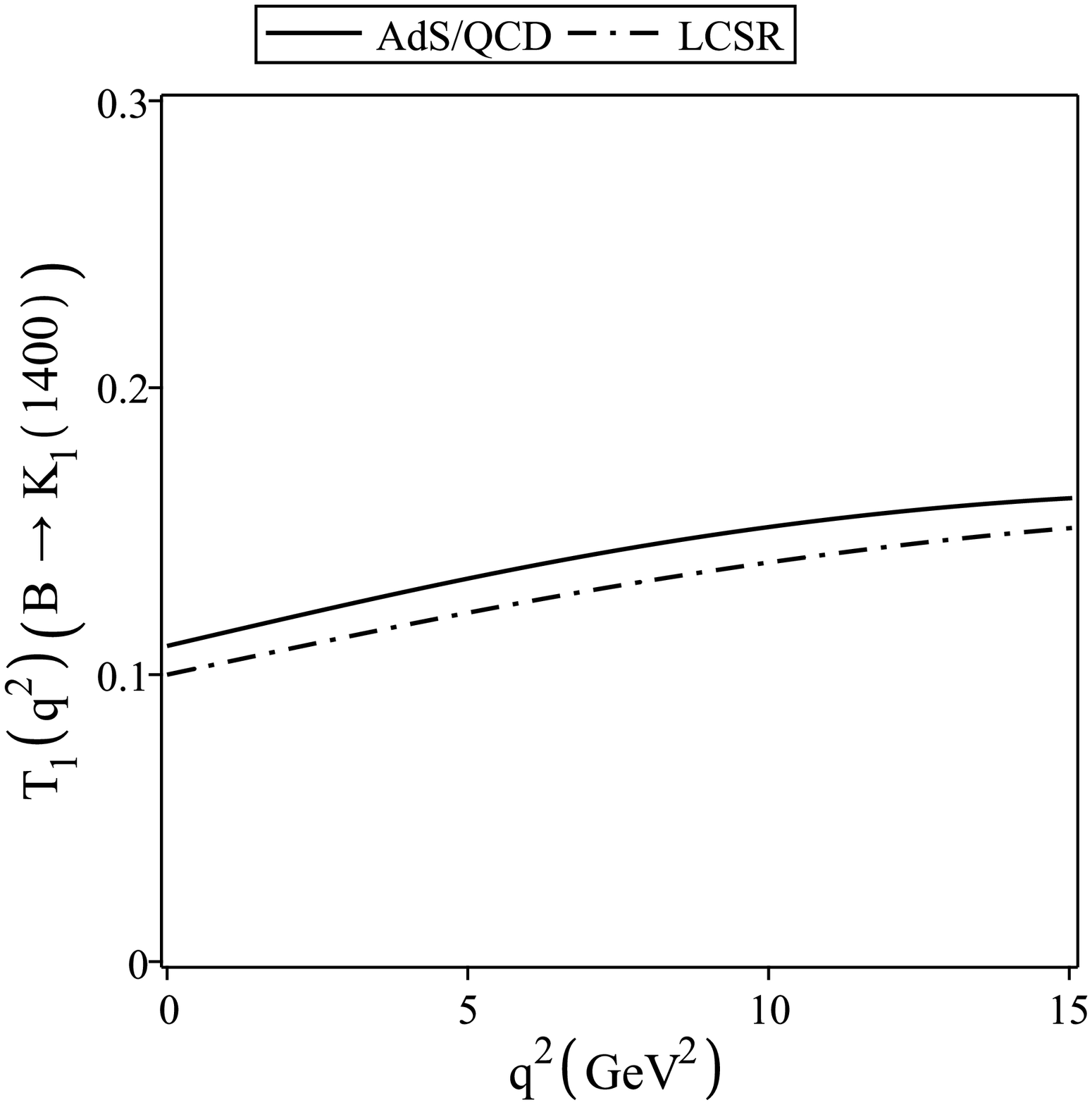}
\caption{The semileptonic form factors $A(q^2)$ and $T_1(q^2)$ for
$B\to K_{1}(1270)$ and $B\to K_{1}(1400)$ transitions on $q^2$ in
the AdS/QCD and LCSR methods. }\label{F303}
\end{figure}

For a better analysis,  we can illustrate the form factors of $B\to
K_{1} (1270)$ and $B\to K_{1} (1400)$ transitions on $q^2$ in the
AdS/QCD and LCSR methods. For instance, Fig. \ref{F303} shows the
form factors $A$ and $T_1$ in $\theta_{K}=-34^{\circ}$ via the
Ads/QCD and LCSR approaches.

\begin{figure}[th]
\includegraphics[width=7.5cm,height=5.5cm]{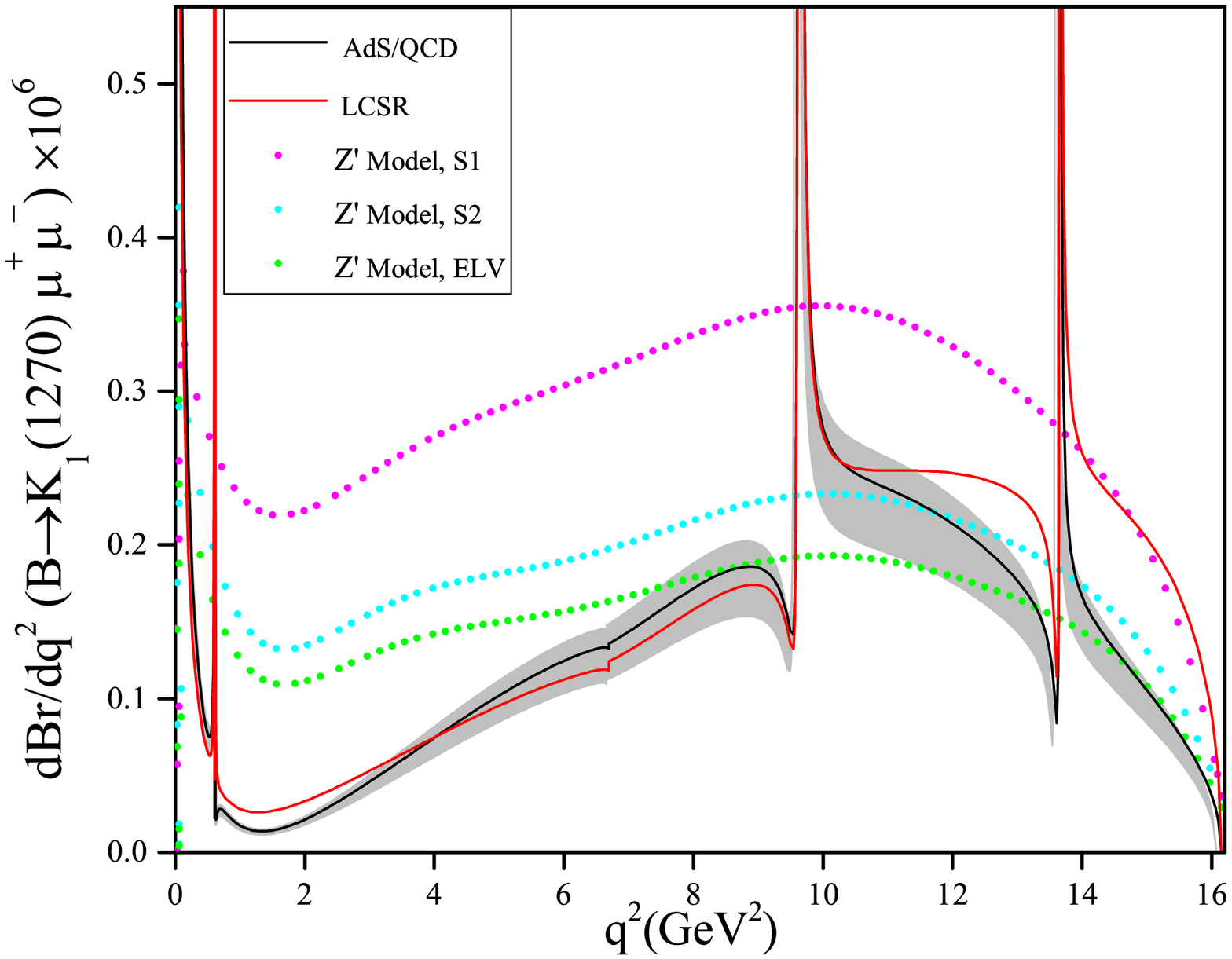}
\includegraphics[width=7.5cm,height=5.5cm]{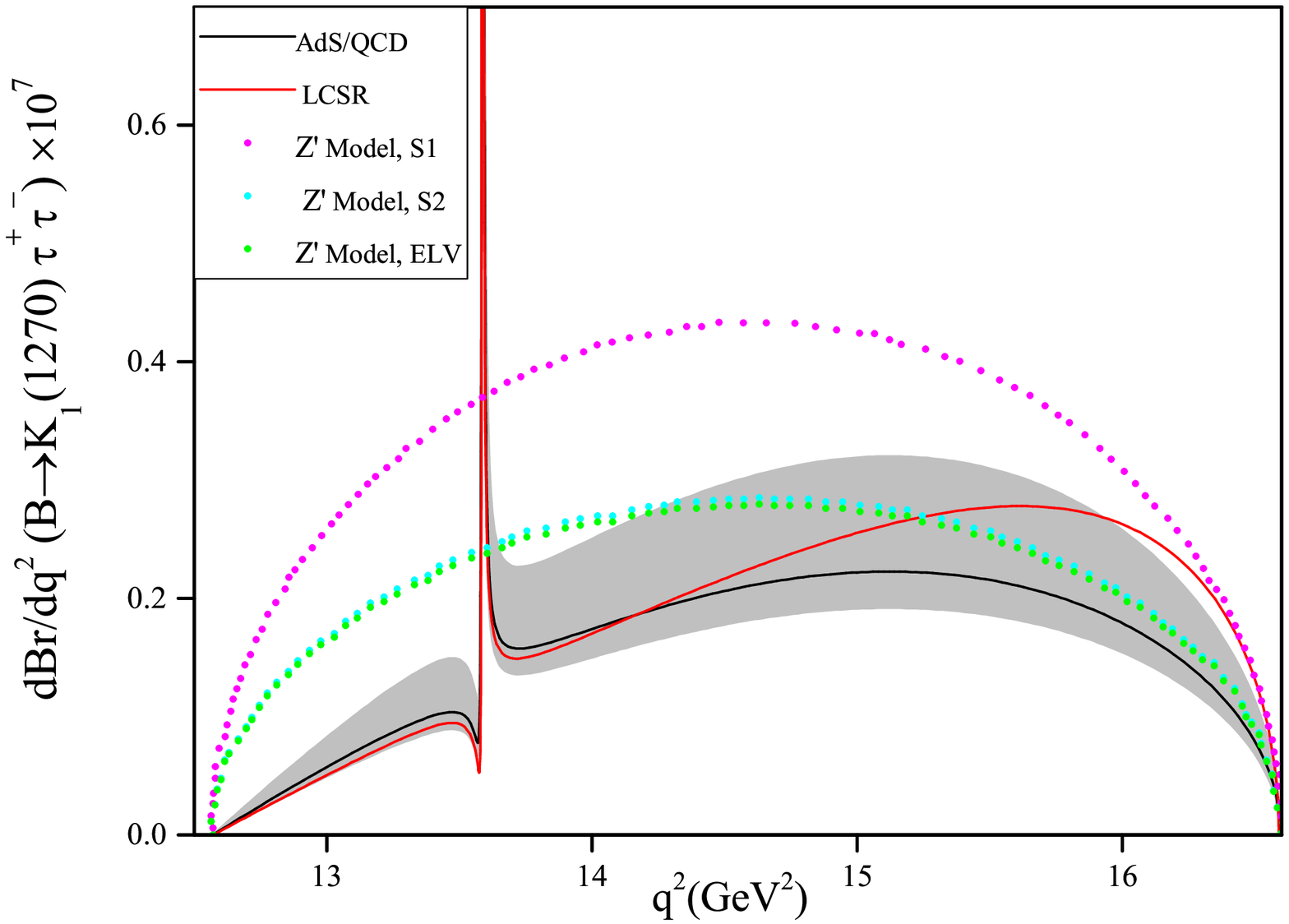}
\includegraphics[width=7.5cm,height=5.5cm]{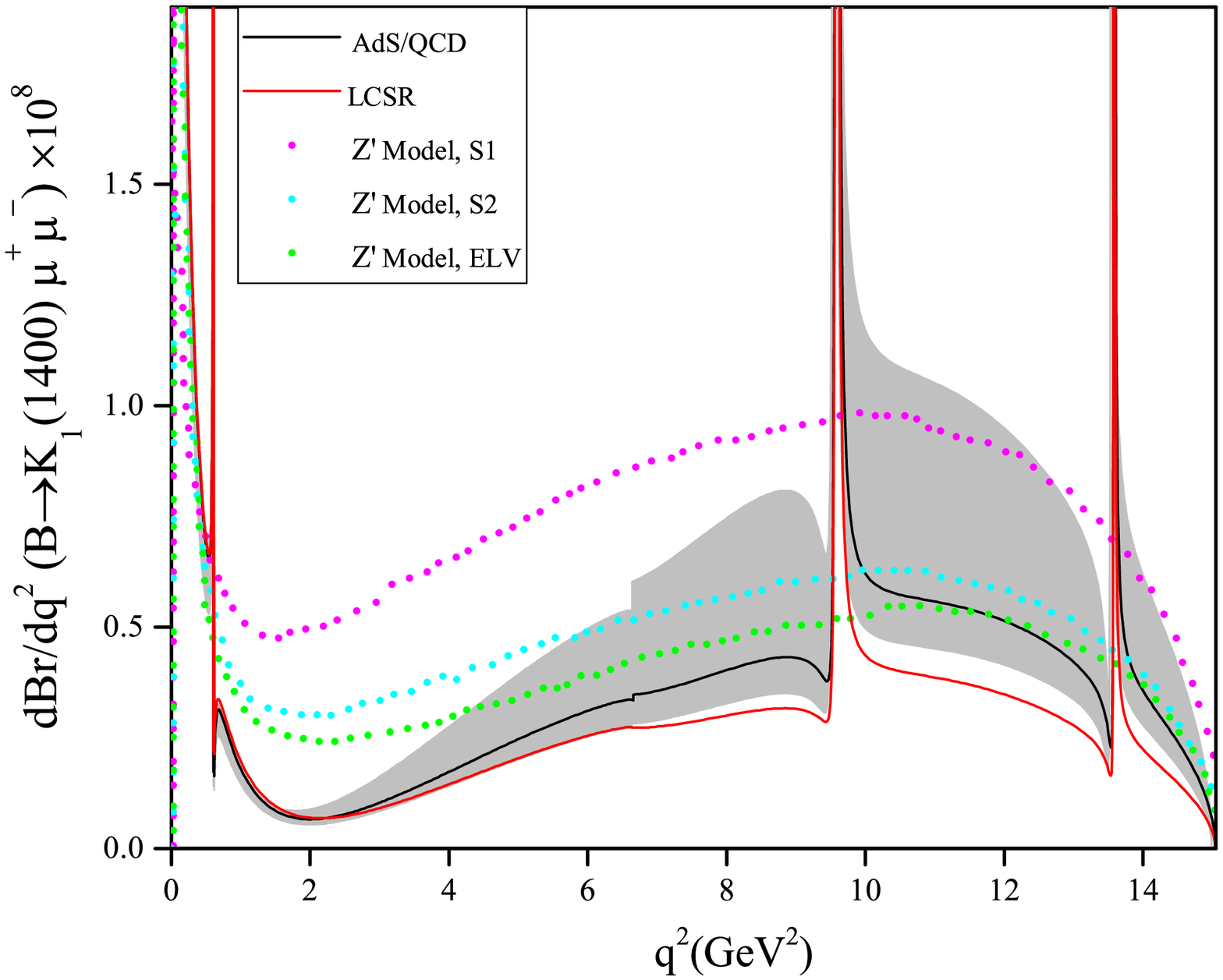}
\includegraphics[width=7.5cm,height=5.5cm]{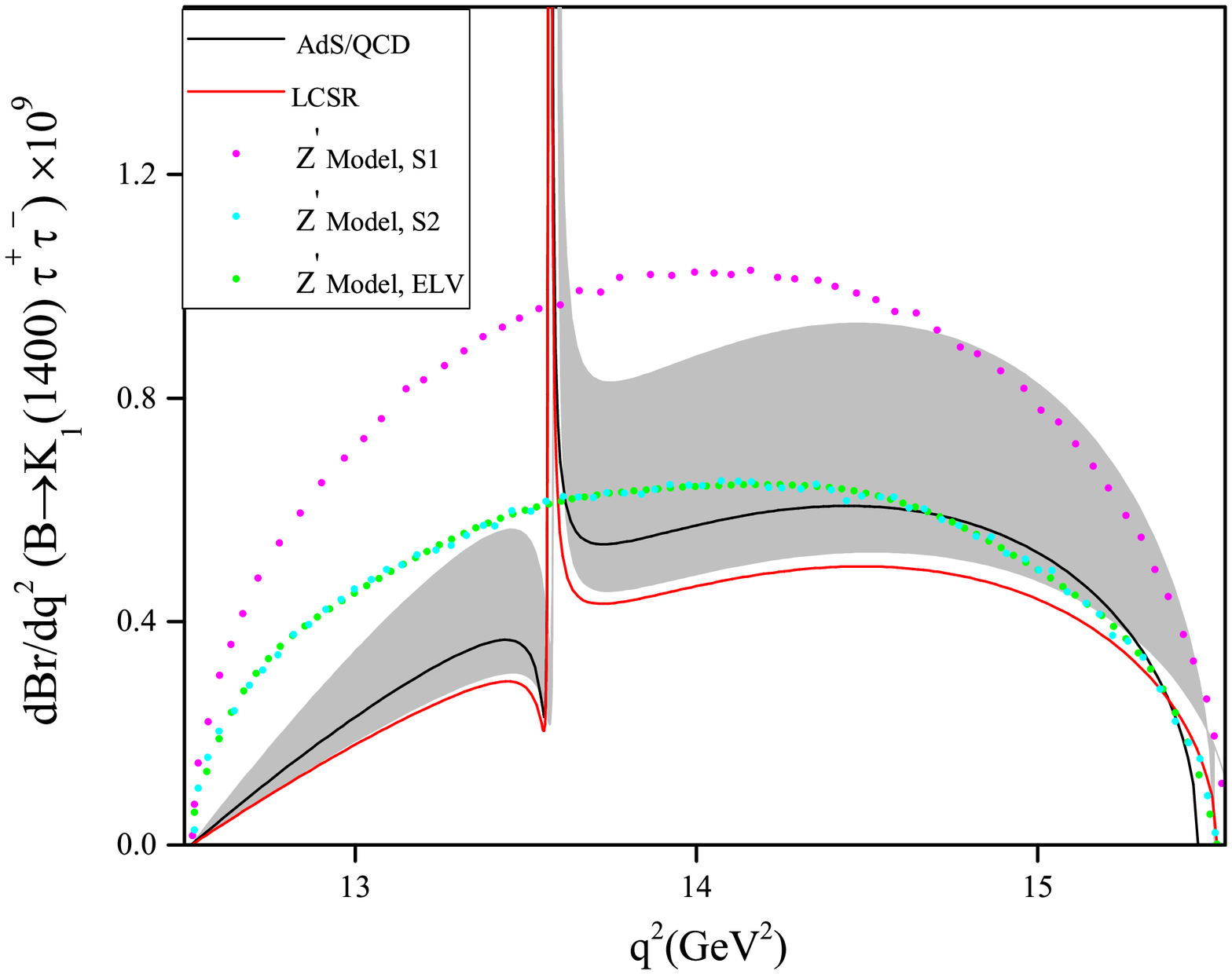}
\caption{The differential branching ratios of the semileptonic $B
\to K_1 \ell^{+}\ell^{-}$ decays  for $ \ell=\mu,\tau$ on $q^2$ via
the AdS/QCD in comparison with the LCSR and $Z'$ model.}\label{F34}
\end{figure}

We would like to plot the differential branching ratios for  $B \to
K_1 \ell^{+} \ell^{-}$ decays with respect to $q^2$. The expression
of double differential decay rate  ${d^2\Gamma}/{dq^2
dcos\theta_{\ell}} $ for  $ B\to K_1$ transitions can be found in
Refs. \cite{Geng2, {Colangelo}}. This expression contains the Wilson
coefficients, the CKM matrix elements, the form factors related to
the fit functions, series of functions and  constants. The numerical
values of the Wilson coefficients are taken from Ref. \cite{Ali}.
The other parameters can be found in Ref. \cite{Colangelo}. After
numerical analysis, the dependency of the differential branching
ratios on $q^2$, by considering the long distance (LD) effects, is
shown in Fig. \ref{F34} in the $\theta_K=-34^{\circ}$. The LD is
associated with real $c \bar c$ resonances in the intermediate
states, i.e., the cascade process $B \to K_{1} J/\psi (\psi')\to K_1
\ell^+ \ell^-$. Fig. \ref{F34} also contains the LCSR and $Z'$ model
predictions \cite{Hua}. It is noted that the results for the
non-universal $Z'$ model are depicted in three sets, considering
only the short distance (SD) effect without the LD effects (for more
details, see Ref. \cite{Hua}). As can be seen, there is some
difference between the predictions of the AdS/QCD and LCSR on one
side and the $Z'$ model, as a method beyond the standard model, on
the other.

Our predictions for the branching ratio values of $B \to K_1
\ell^{+} \ell^{-}$ decays at $\theta_K=-34^{\circ}$ are presented in
Table \ref{T37}.
\begin{table}[th]
\caption{Branching ratio values of  $B \to K_1(1270) \ell^+ \ell^-$
decays at $\theta_K=-34^{\circ}$ in the AdS/QCD correspondence and
LCSR model.} \label{T37}
\begin{ruledtabular}
\begin{tabular}{ccc}
\mbox{Mode}& AdS/QCD & LCSR\\
\hline
\mbox{Br}($B\to K_{1}(1270) \mu^+ \mu^-)\times
10^{6}$&${3.12}\pm {1.14}$& $ {2.91} \pm {1.32}$ \\
\mbox{Br}($B\to K_{1}(1270) \tau^+ \tau^-)\times 10^{7}$&${1.25}\pm
{0.53}$ &${1.07} \pm {0.45}$\\
\mbox{Br}($B\to K_{1}(1400) \mu^+ \mu^-)\times
10^{7}$&${1.13}\pm {0.41}$& ${0.90} \pm {0.33}$ \\
\mbox{Br}($B\to K_{1}(1400) \tau^+ \tau^-)\times 10^{9}$&${1.15}\pm
{0.92}$ &${1.11} \pm {0.90}$
\end{tabular}
\end{ruledtabular}
\end{table}

To evaluate the branching ratio of the non-leptonic $B \to
K_{1}(1270,1400) \gamma$ decays, we use the  exclusive decay width
as \cite{Safir}:
\begin{eqnarray*}\label{eq42}
\Gamma(B \to K_1 \gamma)& = &  \frac{\alpha_{em}\, G_F^2} {32 \pi^4}
m_{b}^5 |V_{tb} V_{ts}^*|^2  |C_{7}(m_b)|^2~ \left (T_1(0)^{B\to
K_1} \right)^2 ~\left (1-{m_{K_1}^2\over m_B^2}\right )^3 \left (1+
{m_{K_1}^2\over m_B^2}\right ).
\end{eqnarray*}
Table \ref{T38} shows our predictions for the branching ratios of
these exclusive non-leptonic decays at $\theta_{K}=-34^{\circ}$. The
AdS/QCD prediction for the branching ratio of the $B \to K_{1}(1270)
\gamma$ decay is larger than the experimental value that is
$(0.43\pm 0.18)\times 10^{-4}$ \cite{Yang}. However, our estimation
has many errors due to the uncertainties in the mixing angle
$\theta_{K}$.
\begin{table}[th]
\caption{AdS/QCD predictions for the branching ratios of  $B \to
K_1(1270,1400) \gamma$ decays in $\theta_{K}=-34^{\circ}$.}
\label{T38}
\begin{ruledtabular}
\begin{tabular}{ccc}
\mbox{Mode} &AdS/QCD & EXP  \cite{Yang} \\
\hline \mbox{Br}($B\to K_{1}(1270) \gamma)\times 10^{4}$& $
{0.71}\pm 0.23$  &  $ {0.43}\pm {0.18}$\\
\mbox{Br}($B\to K_{1}(1400) \gamma)\times 10^{5}$& $
{1.56}\pm{1.04}$  &  $ < 1.44 $
\end{tabular}
\end{ruledtabular}
\end{table}

\begin{figure}[th]
\includegraphics[width=7.5cm,height=5.65cm]{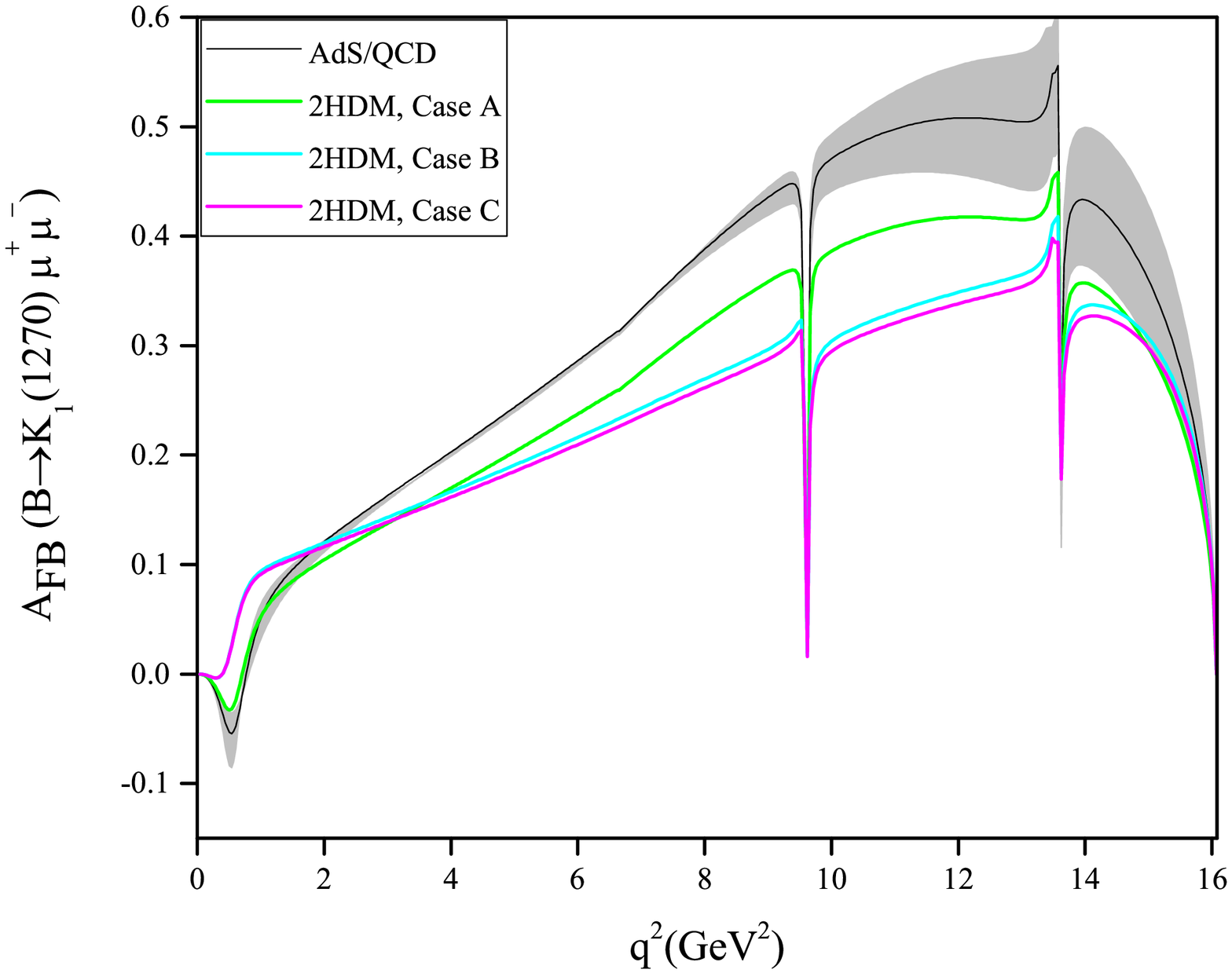}
\includegraphics[width=7.5cm,height=5.5cm]{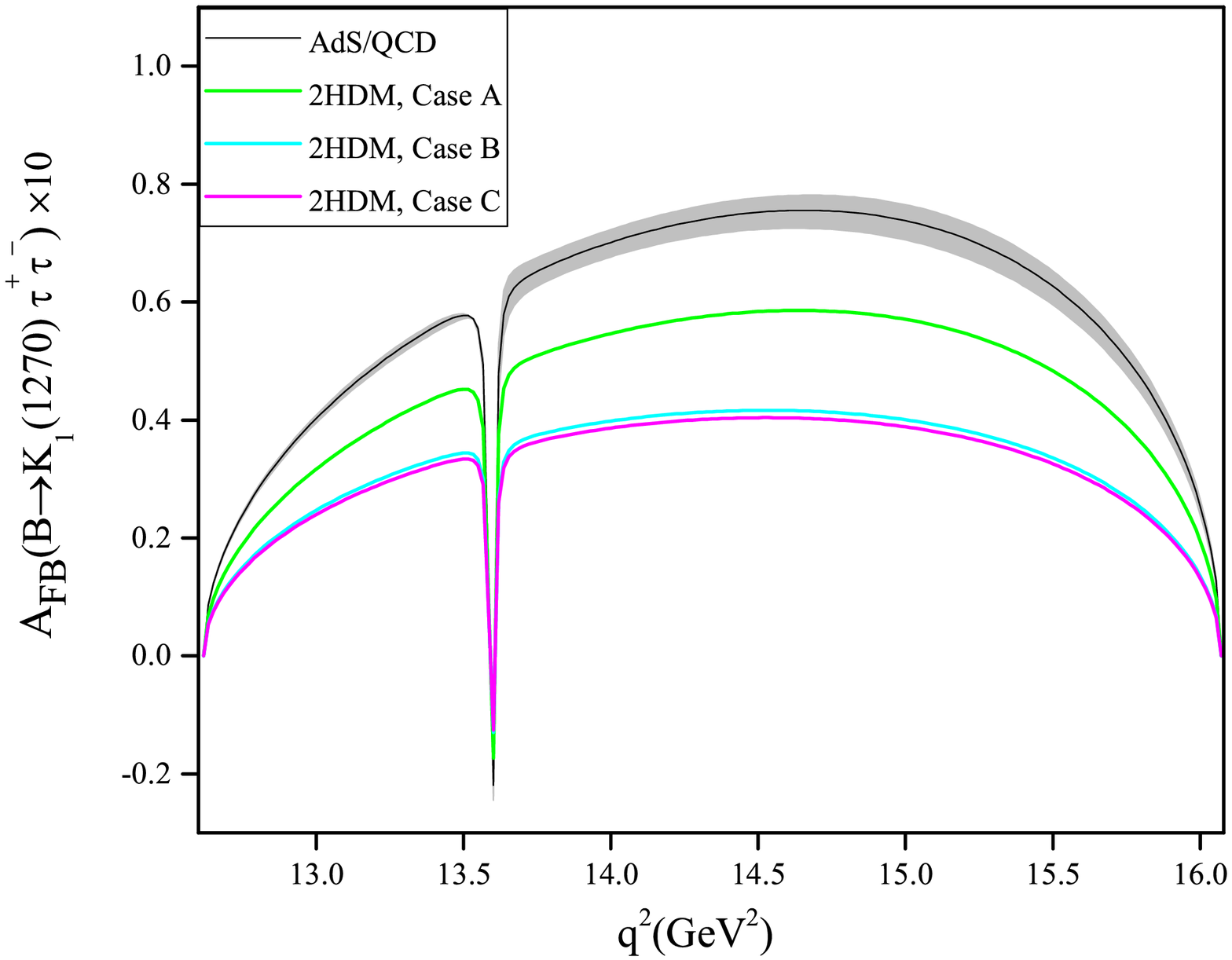}
\includegraphics[width=7.5cm,height=5.5cm]{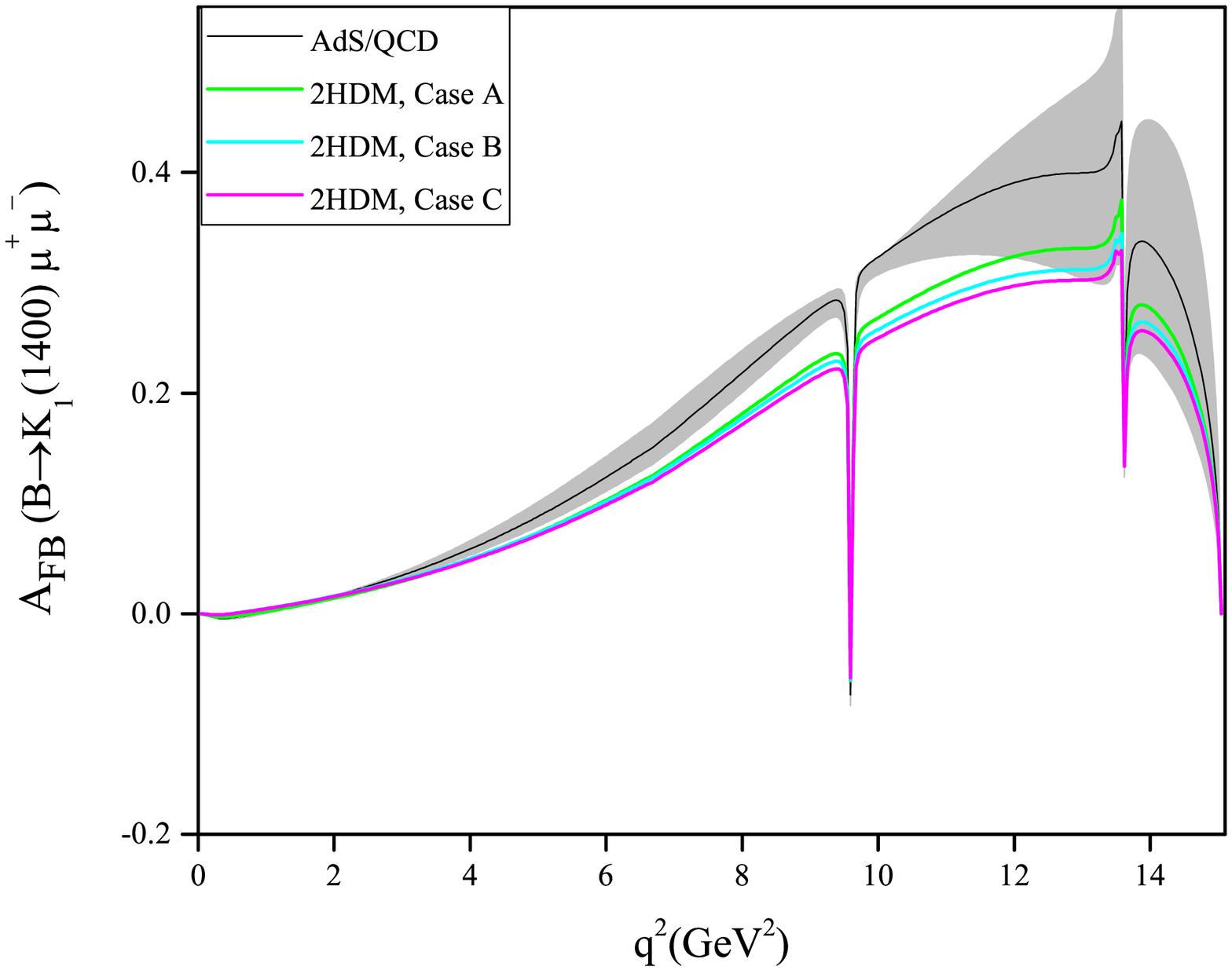}
\includegraphics[width=7.5cm,height=5.5cm]{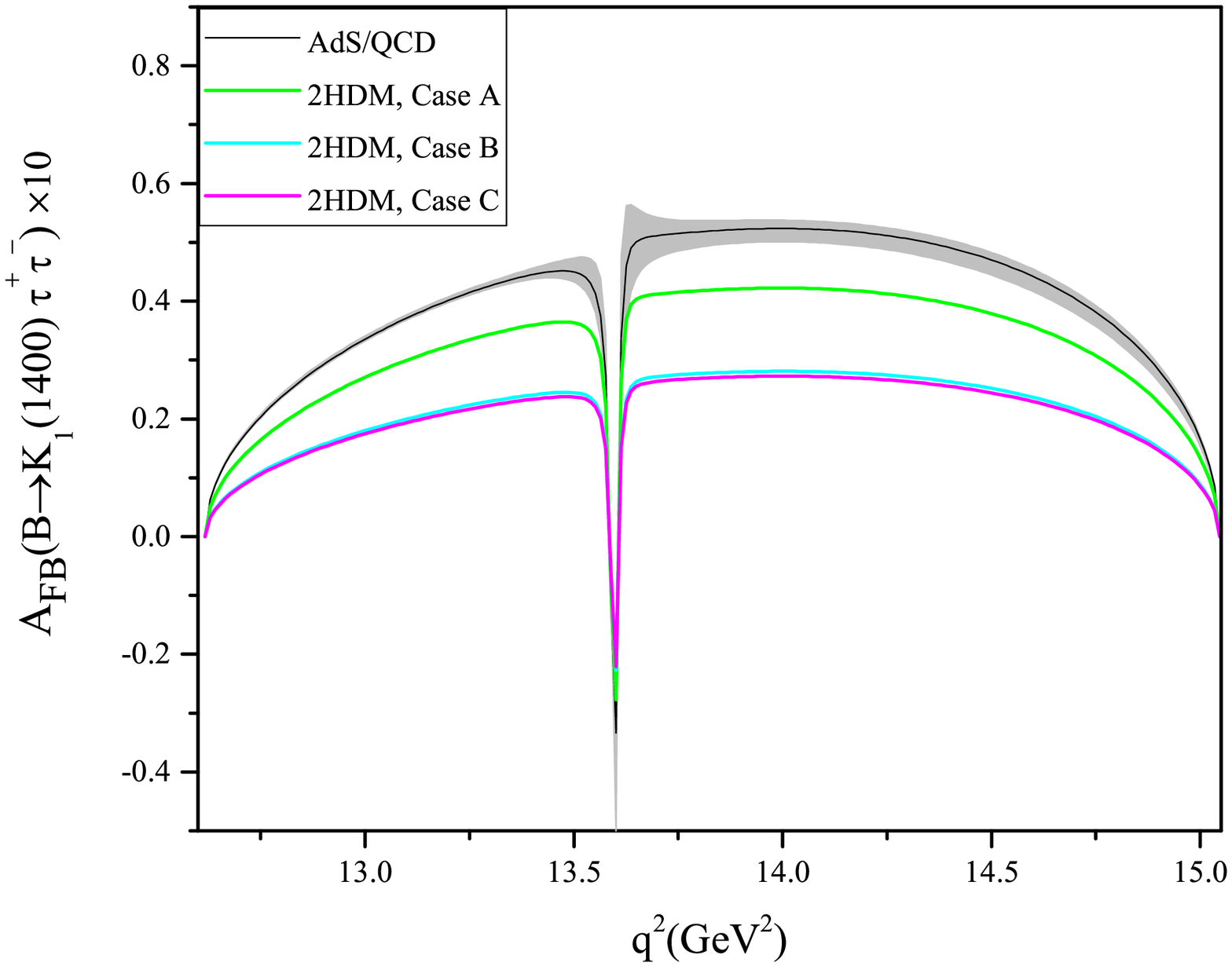}
\caption{ Dependence of the forward-backward asymmetries for $B \to
K_{1} \ell^{+}\ell^{-} (\ell=\mu, \tau)$ decays  on $q^2$ with the
AdS/QCD, LCSR and 2HDM approaches. Gray areas show the errors of the
AdS/QCD correspondence. }\label{F35}
\end{figure}

Finally, we plot dependence of the forward-backward asymmetries,
$A_{FB}$, on $q^2$ for $B\to K_1(1270,1400) \ell^+ \ell^-$ decays,
by considering the LD effects, at $\theta_{K}=-34^{\circ}$ in Fig.
\ref{F35}. Gray regions show the errors of the AdS/QCD
correspondences due to the uncertainties of the input parameters. In
this figure, we also present the behavior of the forward-backward
asymmetries with respect to $q^2$ in the frame work of the 2HDM as a
NP model. To draw the 2HDM diagrams, we insert the AdS/QCD form
factors in the 2HDM formalism for three cases A, B and C related to
$\lambda_{tt}$ and $\lambda_{bb}$ (for more details, see Ref.
\cite{Falahati}) in order to compare the AdS/QCD and 2HDM results.

As can be seen in Fig. \ref{F35}, the forward-backward asymmetries
for $B\to K_1(1270,1400) \tau^+ \tau^-$ transitions are positive for
all values of $q^2$ except in the resonance region. On the other
hand, the 2HDM plots are out of the AdS/QCD predictions and its
errors. Therefor, their investigation in experiments will be a very
efficient tool in establishing a new physics.

In summary, we used the AdS/QCD correspondence as a new remarkable
feature of the light-front holography, to derive the
non-perturbative twist-2 DAs  and decay constants for the pure
axial-vector states, $K_{1A}$ and $K_{1B}$. The holographic DAs for
$K_1 (1270)$ and $K_1(1400)$ mesons were calculated in terms of the
DAs for the aforementioned states. Using the holographic DAs for
$K_1 (1270,1400)$ mesons, we evaluated transition form factors of
the FCNC $B\to K_{1}(1270,1400)\,\ell^{+}\,\ell^{-}$ decays. A
comparison was made between our results and the LCSR predictions for
the twist-2 DAs, decay constants and form factors. We presented our
results for the branching ratio values of the leptonic $B\to
K_{1}(1270,1400)\,\ell^{+}\,\ell^{-}$, $(\ell=\mu, \tau)$, and
non-leptonic $B \to K_1(1270,1400)\gamma$ decays  at the mixing
angle $\theta_{K}=-34^\circ$. The AdS/QCD prediction for the
branching ratio of the $B \to K_1(1270)\gamma$ decay is larger than
the experimental value. Finally, considering the LD effects, we
showed the dependence of the forward-backward asymmetries $A_{FB}$
on $q^2$ for $B\to K_1(1270,1400) \ell^+ \ell^-$ decays at
$\theta_{K}=-34^{\circ}$ in the framework of the AdS/QCD and 2HDM.
Since there was not an overlap between the results of $A_{FB}(B\to
K_1(1270,1400) \tau^+ \tau^-$ from two theories, their experimental
investigation can serve as a crucial test in search of new physics.

\section*{Acknowledgments}
Partial support from the Isfahan university of technology research
council is appreciated.

\clearpage
\appendix

\begin{center}
\section*{\textbf{Appendix:  Expressions for the form factors}}\label{app:form factors}

\end{center}
In this appendix, the explicit expressions for the form factors of
the FCNC $B\to K_{1}(1270,1400)\,\ell^{+}\,\ell^{-} $ decays are
presented.
\begin{eqnarray*}
A(q^{2})&=& \frac{f_{A}
m_{b}}{4\,m_{B}^{2}\,f_{B}}\,(m_{A}-m_{B})\Bigg\{\frac{f_{A}^\perp}{f_{A}}
\int_{u_0}^{1} du~\frac{9\,\Phi^\perp (u)}{u}e^{s(u)}
+\frac{m_{b}}{4m_{A}}\int_{u_0}^{1}du~\frac{{g_\perp^{(v)\prime}(u)}}{u}e^{s(u)}\nonumber\\
&-&\frac{m_{b}}{4m_{A}}\int_{u_0}^{1}du~\frac{{g_\perp^{(v)}(u)}}{u^2}\left[1+\frac{\delta_{1}(u)-8m_{A}^{2}}{M^{2}}\right]e^{s(u)}
+\frac{f_{A}^\perp m_{A}^{2}}{f_{A}}\int_{u_0}^{1}du~\frac{32~\bar{h}{_\parallel^{(t)(ii)}(u)}}{M^2}e^{s(u)}\Bigg\},\nonumber\\
V_{1}(q^{2})&=&
-\frac{m_{b}}{8\,m_{B}^{2}\,f_{B}}\,\frac{f_{A}^\perp}{(m_{B}-m_{A})}\Bigg\{\frac{1}{2}\int_{u_0}^{1}du
~\frac{7\,\Phi^\perp
(u)\,\delta_{1}(u)}{u}e^{s(u)}+2\,m_{A}^{2}\int_{u_0}^{1}du\frac{h_\parallel^{(p)}(u)}{u}
~e^{s(u)}\nonumber\\
&-&3\frac{f_{A}}{f_{A}^\perp}m_{A}m_{b}\int_{u_0}^{1}du~\frac{{g_\perp^{(a)}(u)}}{u}~e^{s(u)}
-8\,m_{A}^{2}\,\int_{u_0}^{1}du~
\frac{\bar{h}{_\parallel^{(t)(ii)}}(u)}{u^{2}}~e^{s(u)}\Bigg\},\nonumber \\
V_{2}(q^{2})&=& -\frac{f_{A}^\perp
m_{b}}{4\,m_{B}^{2}\,f_{B}}\,{(m_{B}-m_{A})}\Bigg\{18\int_{u_0}^{1}du
\frac{\Phi^\perp (u)}{u}e^{s(u)}+\frac{4f_{A}m_{A}m_{B}}{f_{A}^\perp}\int_{u_0}^{1}du~\frac{\phi_{a}(u)}{u^2\,M^2} e^{s(u)} \nonumber\\
&+& 4m_{A}^{2}
\int_{u_0}^{1}du\frac{{h_\parallel^{(p)}(u)}}{u}(1+2u)e^{s(u)}+\frac{16f_{A}}{f_{A}^\perp}  m_{A}m_{b}
 \int_{u_0}^{1}du~\frac{{\Phi}^{{\|}(i)}(u)}{u^{2}M^{2}}~e^{s(u)}
 \nonumber\\
&-&\left.16m_{A}^{2}\int_{u_0}^{1}du
\frac{\bar{h}{_\parallel^{(t)(ii)}}(u)}{u^{2}}\left[\frac{2\,\delta_{3}(u)}{u\,M^4}-\frac{3}{2\,M^2}
 \frac{\delta_{1}(u)}{4u\,M^4}\right]e^{s(u)}\right.\Bigg\},\nonumber\\
V_{0}(q^{2})&=&V_{3}(q^{2})+\frac{m_{b}}{8\,m_{B}^{2}\,f_{B}}\,\frac{f_{A}^\perp
q^2}{m_{A}}\Bigg\{9\int_{u_0}^{1}du \frac{\Phi^\perp
(u)}{u}~e^{s(u)}+2f_{A}\,m_{A}m_{B}\int_{u_0}^{1}du
~\frac{\phi_{a}(u)}{u^2\,M^2}
~e^{s(u)}\nonumber\\
&-&4 m_{A}^{2}
\int_{u_0}^{1}du~\frac{{h_\parallel^{(p)}(u)}}{u}~(1-u)~e^{s(u)}+\frac{16f_{A}}{f_{A}^\perp}m_{A}m_{b}
\int_{u_0}^{1}du~\frac{{\Phi^{{\|}(i)}}(u)}{u^{2}M^{2}}~e^{s(u)}
\nonumber\\
&+&\left.8m_{A}^{2}\int_{u_0}^{1}du~
\frac{\bar{h}_{\parallel}^{(t)(ii)}(u)}{u^{2}}\left[\frac{2\,\delta_{3}(u)}{u\,M^4}
-\frac{1}{M^2}+(1-u)\,(-\frac{1}{M^2}
+\frac{\delta_{1}(u)}{2u\,M^4})\right]~e^{s(u)}\right.\Bigg\},\nonumber\\
T_{1}(q^{2})&=&-\frac{f_{A}
m_{b}}{8\,m_{B}^{2}\,f_{B}}\Bigg\{m_{b}(\frac{f_{A}^\perp}{f_{A}}+8)\int_{u_0}^{1}du
\frac{\Phi^\perp (u)}{u}~e^{s(u)}
-3m_{A}\int_{u_0}^{1}du~g_{\perp}^{(a)}(u)~e^{s(u)}\nonumber\\
&+&4m_{A}
\int_{u_0}^{1}du~\frac{\phi_{a}(u)}{u}~e^{s(u)}-\frac{f_{A}}{
m_{A}}
\int_{u_0}^{1}du~\frac{{g_\perp^{(v)\prime}(u)\,\delta_{5}(u)}}{u}~e^{s(u)}
-\frac{m_{A}}{8}\int_{u_0}^{1}du~\frac{{g_\perp^{(v)\prime}(u)}}{u}\nonumber\\
&\times &\left[7-\frac{\delta_{5}(u)(8u-1)}{M^2}+
\frac{u\delta_{2}(u)-\delta_{4}(u)}{2\,m_{A}^2}\right] ~e^{s(u)}
+4m_{A}\int_{u_0}^{1}du\frac{{\Phi^{\|(i)}}(u)}{u}~e^{s(u)}-\frac{16f_{A}^\perp}{f_{A}}
\nonumber\\
&\times&\left.
m_{A}^{2}\,m_b\int_{u_0}^{1}du
\frac{\bar{h}{_\parallel^{(t)(ii)}}(u)}{u\,M^2}~e^{s(u)}
\right.\Bigg\},\nonumber\\
T_{2}(q^{2})&=&\frac{m_{b}}{m_{B}^{2}\,f_{B}}\frac{f_{A}}{m_{A}^2-m_{B}^2}\Bigg\{\frac{m_{b}f_{A}^\perp}
{f_{A}}\int_{u_0}^{1}du ~\frac{\Phi^\perp
(u)\delta_{1}(u)}{u}~e^{s(u)}
+\frac{1}{2}m_{A}\int_{u_0}^{1}du \frac{~g_{\perp}^{(a)}(u)}{u}\nonumber\\
&\times&
\left[\delta_{1}(u)+4\,\delta_{5}(u)\right]~e^{s(u)}-\frac{1}{16}
m_{A}
\int_{u_0}^{1}du~\frac{{g_\perp^{(v)\prime}(u)\,\delta_{2}(u)}}{u}~e^{s(u)}
+\frac{1}{2}{m_{A}}\int_{u_0}^{1}du~\frac{\phi_{a}(u)}{u}\nonumber\\
&\times&
\delta_{1}(u)e^{s(u)}+{m_{A}}\int_{u_0}^{1}du~\frac{g_\perp^{(v)}(u)}{u^2}\left[\delta_{6}(u)+
\frac{\delta_{1}(u)\delta_{5}(u)}{M^2}+\frac{u \delta_{2}(u)}{2}
\Bigg(1+\frac{\delta_{3}(u)
}{u\,M^2}+\frac{\delta_{7}(u)}{u}\Bigg) \right.\nonumber\\
 &+& \left.u
\Bigg(m_{A}^2-2\delta_{1}(u)+\frac{\delta_{4}(u)}{2}+\frac{\delta_{5}(u)\delta_{1}(u)}{u\,M^2}\Bigg)\right]e^{s(u)}
-2m_{A}\int_{u_0}^{1}du~\frac{{\Phi_{\|}^{(i)}(u)\delta_{2}(u)}}{u}e^{s(u)}\nonumber\\
&-&8\frac{f_{A}^\perp}{f_{A}}\,m_{A}^{2}\,m_b\int_{u_0}^{1}du~\frac{\bar{h}{_\parallel^{(t)(ii)}}(u)}{u^2}
\left[1+\frac{\delta_{2}(u)}{M^2}\right]e^{s(u)}\Bigg\},\nonumber\\
\end{eqnarray*}
\begin{eqnarray*}
T_{3}(q^{2}) &=& -\frac{f_{A} m_{b}}{4\,m_{B}^{2}\,f_{B}}
\Bigg\{\frac{8f_{A}^\perp}{f_{A}}m_{b}\int_{u_0}^{1}du
~\frac{\Phi^\perp (u)}{u}~e^{s(u)} -4m_{A}\int_{u_0}^{1}du
\frac{~g_{\perp}^{(a)}(u)}{u}~e^{s(u)}
-\frac{1}{4\, m_{A}}\nonumber\\
&\times&\left.
\int_{u_0}^{1}du~\frac{{g_\perp^{(v)\prime}(u)}}{u}~\left[\frac{7}{2}\delta_{2}(u)+m_{A}^2-
\frac{\delta_{1}(u)}{4\,u} \right]e^{s(u)}
-m_{A}\int_{u_0}^{1}du~\frac{\phi_{a}(u)}{u^2}[\frac{u\,\delta_{1}(u)+2\,\delta_{2}(u)}{M^2}\right.\nonumber\\
&+&\left.\frac{1}{u}]\,e^{s(u)}-
4m_{A}\int_{u_0}^{1}du~\frac{{\Phi}^{\|(i)}(u)}{u^2}\left[\frac{u\,\delta_{5}(u)-\delta_{2}(u)}{M^2}
-1\right]e^{s(u)}-\frac{1}{4m_{A}}\int_{u_0}^{1}du\frac{{g_\perp^{(v)}(u)}}{u} \right.\nonumber\\
&\times&\left.
\left[\frac{3\,\delta_{1}(u)}{u\,M^2}-\frac{\delta_{1}(u)}{m_{a_1}^2}+
\frac{5\,\delta_{5}(u)-7\,\delta_{3}(u)}{M^2}+\frac{\delta_{2}(u)}{M^2}-\frac{\delta_{1}(u)^{2}}
{m_{a_1}^2\,M^2}\right]e^{s(u)}+\frac{16f_{A}^\perp}{f_{A}}\,m_{A}^{2}\,m_b\right.\nonumber\\
&\times&\left.
\int_{u_0}^{1}du~
\frac{\bar{h}{_\parallel^{(t)(ii)}}(u)}{u^2\,M^2}\left[8+\frac{2}{u}+\frac{\delta_{2}(u)}{u\,M^2}\right]e^{s(u)}\right.\Bigg\},
\end{eqnarray*}
where
\begin{eqnarray*}
u_{0} &=&\frac{1}{2m_{A}^2} \left[\sqrt{(s_0-m_{A}^2-q^2)^2 +4 m_{A}^2 (m_b^2-q^2)} -\left(s_0-m_{A}^2-q^2\right)\right],\nonumber\\
s(u)&=&-\frac{1}{uM^2}\left[m_b^2+u\bar{u}m_{A}^2-\bar{u}q^2\right]+\frac{m_B^2}{M^2},\nonumber\\
\delta_{1}(u)&=& m_{A}^2(u+2)+\frac{m_{b}^2}{u}+\frac{q^2}{u},\,\,\,\,\,\,\,
 \delta_{2}(u)=u\,m_{A}^2-\frac{m_{b}^2}{u}+q^2\, \frac{u-\bar{u}}{u},\nonumber\\
\delta_{3}(u)&=& \frac{m_{b}^2}{u}-2q^2\, \frac{\bar{u}}{u},\,\,\,\,\,\,\,\,\,\,\,\,\,\,
\,\,\,\,\,\,\,\,\,\,\,\,\,\,\,\,\,\,\,\,\,\,\delta_{4}(u)= 2\,m_{A}^2(u+1)+2q^2,\nonumber\\
\delta_{5}(u)&=&u\,m_{A}^2-\frac{m_{b}^2}{u}+\frac{q^2(u-2)}{u},\,\,\,\,
\,\,\delta_{6}(u)= 2\,m_{A}^2(u+1)+q^2\frac{\bar{u}}{u},\nonumber\\
\delta_{7}(u)&=&-2\frac{m_{b}^2}{u}+\frac{q^2}{u},\,\,\,\,\,\,\,\,\,\,\,\,\,\,\,\,\,\,
\,\,\,\,\,\,\,\,\,\,\,\,\,\,\,\,\,{f}^{(i)}(u)\equiv\int_0^u f(v) dv,\nonumber\\
{f}^{(ii)}(u)&\equiv&\int_0^u dv\int_0^v d\omega f(\omega),\,\,\,\,\,\,\,\,\,\,\,\,\,\,\,\,\,\,
\,\,\,\,\,\,\,\,\,\,\,\,\bar h_\parallel^{(t)} = h_\parallel^{(t)}- \frac{1}{2} \Phi^{\perp}(u),\nonumber\\
\phi_a(u)&=& \int_0^u \left[\Phi^\parallel - g_\perp^{(a)} (v)\right]dv.\nonumber\\
\end{eqnarray*}

\end{document}